\documentstyle[12pt,psfig,subeqnarray]{article}
\def\erf{\mbox{erf}}
\def\Ac{{\cal A}}
\def\Bc{{\cal B}}
\def\Rc{{\cal R}}

\title{Asymptotics of Reaction-Diffusion Fronts with One
Static and One Diffusing Reactant}

\author{Martin Z. Bazant \\ {\small {\it
Department of Mathematics, Massachusetts Institute of Technology,
Cambridge, MA 02139}} \\
\ \\
H. A. Stone \\
{\small {\it Division of Engineering and Applied
Sciences, Harvard University, Cambridge, MA 02138}}
}

\date{June 8, 2000}

\begin{document}
\maketitle

\begin{abstract}
The long-time behavior of a reaction-diffusion front between one
static ({\it e.g.} porous solid) reactant A and one initially
separated diffusing reactant B is analyzed for the mean-field
reaction-rate density $R(\rho_A,\rho_B) = k\rho_A^m\rho_B^n$. A
uniformly valid asymptotic approximation is constructed from matched 
self-similar solutions in a \lq\lq reaction front'' (of width $w \sim
t^\alpha$ where $R \sim t^{\beta}$ enters the dominant balance) and a 
\lq\lq diffusion layer'' (of width $W \sim t^{1/2}$ where $R$ is
negligible).  The limiting solution exists if and only if $m, n \geq
1$, in which case the scaling exponents are uniquely given by $\alpha
= (m-1)/2(m+1)$ and $\beta = m/(m+1)$. In the diffusion layer, the
common {\it ad hoc} approximation of neglecting reactions is given
mathematical justification, and the exact transient decay of the
reaction rate is derived. The physical effects of higher-order
kinetics ($m, n > 1$), such as the broadening of the reaction front
and the slowing of transients, are also discussed.
\end{abstract}

\noindent {\it PACS:} 05.40+j, 82.20.-w, 02.30.Jr
\vskip 12pt
\noindent {\it Keywords:} Reaction kinetics; diffusion; 
partial differential equations; asymptotic
analysis; similarity solutions.
\vskip 12pt
\noindent {\it Corresponding Author:} Martin
Z. Bazant, Deptartment of Mathematics, Building 2-363B, Massachusetts
Institute of Technology, Cambridge, MA 02139-4307;
bazant@math.mit.edu; phone: (617) 253-1713; fax: (617) 253-8911.
\vskip 24pt

\section{Introduction}

In the decade that has passed since the pioneering analytical study of
G\'alfi and R\'acz~\cite{galfi}, there has emerged a substantial body
of research devoted to experimental~\cite{expt,leger},
computational~\cite{jiang,cornell,cornell97,araujo,havlin,kozatait,tait2}
and analytical~\cite{schenkel,vanbaalen,koza1,koza2,peletier96,peletier99} studies of
reaction-diffusion systems with two initially separated, diffusing
species A and B reacting to produce an inert product C according to
the chemical formula
\begin{equation}
m^\prime\mbox{A} + n^\prime\mbox{B}
\rightarrow \mbox{C (inert)} ,
\end{equation}
where $m^\prime$ and $n^\prime$, the stoichiometric coefficients, are
positive integers. Theoretical studies have focused almost exclusively
on the \lq\lq one-dimensional'' case of an infinite, flat reaction
front between two regions of homogeneous composition of either A or B
(see Fig.~\ref{fig:cartoon}). This idealized situation is believed to
capture much of the essential physics of reaction fronts commonly
observed in various chemical~\cite{grindrod,alberty} and
biological~\cite{murray77,murray93} systems.

The standard continuum model for such a one-dimensional reaction front
involves a pair of nonlinear partial differential
equations~\cite{galfi,cornell,schenkel,vanbaalen,koza1}
\begin{subeqnarray}
\frac{\partial \rho_A}{\partial T} & = & D_A \frac{\partial^2 \rho_A}{\partial
X^2} - m^\prime R(\rho_A,\rho_B) \slabel{eq:coup1} \\
\frac{\partial \rho_B}{\partial T} & = & D_B \frac{\partial^2 \rho_B}{\partial
X^2} - n^\prime R(\rho_A,\rho_B) \slabel{eq:coup2} ,
\label{eq:coup}
\end{subeqnarray}
subject to the boundary conditions
\begin{equation}
\rho_A(-\infty,T) = 0, \ \ \rho_A(\infty,T) = \rho_A^o, \ \
\rho_B(-\infty,T) = \rho_B^o, \ \ \rho_B(\infty,T) = 0
\label{eq:rhoBC}
\end{equation}
and the initial conditions
\begin{equation}
\rho_A(X,0) = \rho_A^o H(X) , \ \
\rho_B(X,0) = \rho_B^o H(-X) \label{eq:rhoIC}
\end{equation}
where $\rho_A(X,T)$ and $\rho_B(X,T)$ are the concentrations, $D_A$
and $D_B$ the diffusion coefficients of A and B, respectively,
$\rho_A^o > 0$ and $\rho_B^o > 0$ are constants, $H(X)$ is the
Heaviside unit step function and $R(\rho_A,\rho_B)$ is the reaction
rate density for production of species C. 
(Note that upper-case letters denote quantities with dimensions,
e.g. $X$ and $T$ for space and time, respectively.  Lower-case letters
for the corresponding dimensionless quantities are introduced in
section~\ref{sec:dim}.)
The reactants are completely separated at first according to
(\ref{eq:rhoIC}), but for $T > 0$ they diffuse together and react,
which decreases the concentrations wherever $\rho_A(X,T)\rho_B(X,T) >
0$. Diffusion acts to replenish any depleted regions. As a result the
system develops a localized, moving region, the \lq\lq reaction
front,'' where the reaction rate $R(\rho_A,\rho_B)$ is greatest and
which is fed by diffusion from the distant particle reservoirs
described by the boundary conditions. The dynamics of this reaction
front are described by the long-time asymptotics of the nonlinear
initial-boundary-value problem (\ref{eq:coup})--(\ref{eq:rhoIC}).

The nonlinear reaction term $R(\rho_A,\rho_B)$ is usually assumed to
have the form of a power law
\begin{equation}
R(\rho_A,\rho_B) = k \rho_A^m \rho_B^n ,
\label{eq:Rpower}
\end{equation}
where $k$ is a rate constant, and $m$ and $n$ are respectively the
\lq\lq kinetic orders'' of A and B in the reaction~\cite{alberty}. For
a one-step reaction with sufficient mixing (see below) $m=m^\prime$
and $n=n^\prime$, but for more complex, multi-step reactions $m$ and
$n$ are determined by the stoichiometric coefficients of the (often
unknown) rate-limiting step. Although $m$ and $n$ are usually taken to
be positive integers, non-integer values of $m$ and $n$ can arise in
certain situations~\cite{alberty}. We will see that a well-defined
reaction front exists for any real numbers $m, n \geq 1$, but not for
$m<1$ or $n<1$.

Technically, by assuming in (\ref{eq:coup}) that the reaction rate $R$
depends only on the average local concentrations (and not on any
fluctuations or many-body effects) we have made the \lq\lq mean-field
approximation''~\cite{cardy}. In low-dimensional systems, such as ion
channels ($d=1$) or catalytic surfaces ($d=2$), the mean-field
approximation can break down because the reacting particles cannot mix
efficiently enough, but as the dimension of the system is increased
above a certain \lq\lq upper critical dimension'' $d_c$, such
statistical anomalies disappear.  For two diffusing reactants with a
simple one-step reaction it is known~\cite{cornell,havlin,cardy} that
$d_c = 2/(m+n-1)$. Since $d_c
\leq 2$ for $m,n \geq 1$ the mean-field approximation should be
perfectly valid in the usual case $d=3$, which is consistent with
experimental findings~\cite{expt}.

In contrast to the case of two diffusing reactants described above,
relatively little is
known~\cite{jiang,havlin,koza2,peletier96,peletier99} about the case
of one diffusing reactant ($D_A > 0$) and one static reactant ($D_B =
0$).  This situation, depicted schematically in
Fig.~\ref{fig:cartoon}, describes the corrosion of a porous solid B
saturated with a fluid solvent and exposed to an initially separated
colloidal reactant A, as shown in recent electrochemical experiments
(described below) ~\cite{leger}. Jiang and Ebner~\cite{jiang} first
pointed out (for $m=n=1$) that setting $D_B=0$ in (\ref{eq:coup}) is a
non-trivial, i.e. singular, limit leading to different long-time
behavior than in the case of $D_B > 0$ (no matter how small), which
they explained with simple scaling arguments supported by Monte Carlo
computer simulations. For an analytical description of such
one-dimensional diffusion with one static reactant, we adopt the
power-law form of the reaction term and study the coupled equations
\begin{subeqnarray}
\frac{\partial \rho_A}{\partial T} & = & D_A \frac{\partial^2
  \rho_A}{\partial X^2} - m^\prime k \rho_{A}^m\rho_{B}^n \slabel{eq:rhoA} \\
\frac{\partial \rho_{B}}{\partial T} & = & - n^\prime k
\rho_{A}^m\rho_{B}^n .
\slabel{eq:rhoB}
\label{eq:rho}
\end{subeqnarray}
In the simplest case $m=n=1$, the initial-boundary-value problem
(\ref{eq:rhoBC})--(\ref{eq:rho}) has been solved numerically by Havlin
et al.~\cite{havlin} and analyzed in the limit of ``long times'' $T
\rightarrow \infty$ by Koza~\cite{koza2}, using various asymptotic
approximations introduced by G\'alfi and R\'acz~\cite{galfi}. Rigorous
analysis has been reported in the analogous limit of ``fast
reactions'' $k\rightarrow\infty$ by Hilhorst et al.~\cite{peletier96},
but these authors only address the behavior at the diffusive length
scale $X \propto \sqrt{T}$ (see section~\ref{sec:diff} below) and do
not consider the structure of the reaction front studied by
Koza~\cite{koza2}, which is of primary interest here.  Hilhorst et
al. have also recently considered the effect of a more general
reaction term at the diffusive scale~\cite{peletier99}, but the
present work appears to be the first to analyze the nontrivial effect
of changing reaction orders at the reactive length scale (see below)
in the general case ($m,n\geq 1$) with one static reactant.

The relevance of (\ref{eq:rho}) for a given porous-solid corrosion
system rests on several key assumptions that are less obviously
satisfied {\it a priori} than in the case of two diffusing reactants.
First, the solid matrix containing the static reactant B must be
sufficiently porous that the moving reactant A can diffuse freely to
the exposed surfaces with an effective diffusion constant (averaged
over many pores) comparable to that in the bulk solvent. The
concentration of A must also be dilute enough that $D_A$ is constant.
Another reason that the concentrations of A and B must be dilute is
that the inert product C must be created in small enough quantities
that its presence does not affect the reaction dynamics (e.g. by
inhibiting diffusion or initiating convection). Finally, one might
worry about the breakdown of the mean-field approximation since the
(possibly fractal) pore structure may influence statistical
averaging. For example, it is known that fluctuations alter the
reaction-front dynamics when the diffusion is confined to a
percolating cluster in two dimensions~\cite{havlin}. In spite of these
concerns, however, the one-dimensional mean-field model (\ref{eq:rho})
can in fact describe certain corrosion systems.

An important motivation for the present analytical study is afforded
by the recent experiments of L\'eger {\it et al.}~\cite{leger}, which
are the first to examine in detail the case of one static and one
diffusing reactant. These experiments involve the corrosion of
ramified copper electrodeposits exposed to a cupric chloride
electrolyte to produce cuprous chloride crystallites via the reaction
\begin{equation}
\mbox{CuCl$_2$ (aq) \ + \ Cu (solid) \ $\rightarrow$ \
2 CuCl (solid)} \label{eq:reaction}
\end{equation}
immediately following electrodeposition. It is found that the
long-time behavior of (\ref{eq:rho}) with $m=1$ matches the
experimentally observed front speed and concentration profile of
diffusing reactant (CuCl$_2$) rather well, in spite of the complex
fractal geometry of the electrodeposits and the presence of the inert
product (CuCl)~\cite{leger}. Since the reaction rate and the
concentration of the static species (Cu) are not directly measured,
however, the interpretation of these kinds of corrosion experiments
can be aided by the analysis presented here of the mean-field model
with $m, n \geq 1$.

There is an extensive mathematical
literature~\cite{grindrod,murray77,murray93,gmira,barenblatt96} on the
subject of single reaction-diffusion equations of the general form
\begin{equation}
\frac{\partial \rho}{\partial T} = D \frac{\partial^2 \rho}{\partial
X^2} - f(\rho) \label{eq:single}
\end{equation}
which arise in many applications (e.g. chemical reactions, combustion
and population dynamics).
A common theme in these studies is the appearance of two distinct
(time-dependent) length scales in the intermediate asymptotic regime
($t \rightarrow \infty$) which correspond to either ``weakly nonlinear
behavior'', where it has been established in many cases that the
reaction term is negligible and the dynamics are purely diffusive, or
``strongly nonlinear behavior'', where the reaction and diffusion
terms balance (in the nomenclature of Gmira and
Veron~\cite{gmira}). This separation of scales also arises in coupled
systems of reaction-diffusion equations like (\ref{eq:coup}), but
owing to their greater complexity, much less rigorous analysis has
been reported.
In the case of two diffusing reactants, G\'alfi and
R\'acz~\cite{galfi} pointed out that if the diffusion constants are
the same, $D_A = D_B$, then the difference in concentrations $\rho_A -
\rho_B$ obeys a pure diffusion equation which can be easily
integrated, thereby reducing the coupled system (\ref{eq:coup}) to a
single equation with the form of (\ref{eq:single}). Another
simplification occurs if also $\rho_A^o = \rho_B^o$ in which case the
reaction front is perfectly symmetric and does not move. In this
simplified case with $m=n=1$, Schenkel et al.~\cite{schenkel} were
able to prove that the asymptotic solution of G\'alfi and
R\'acz~\cite{galfi}, which combines different approximations at the
diffusive and reactive scales, is approached uniformly as $T
\rightarrow \infty$ starting from the initial conditions of
(\ref{eq:rhoIC}), and they also reported rigorous bounds on the
transient decay to the asymptotic solution.  Recently, van Baalen et
al.~\cite{vanbaalen} have extended this analysis to the case of
symmetric, high-order reactions $m=n>3$, where the reaction-front
scaling is altered.

The analyses of Refs.~\cite{schenkel}--\cite{vanbaalen} represent an
important contribution because, at least in the case $D_A=D_B$,
$\rho_A^o=\rho_B^o$ and $m=n$, they provide a rigorous mathematical
justification for various {\it ad hoc} assumptions introduced by
G\'alfi and R\'acz~\cite{galfi,koza1} to describe the local structure
of the reaction front which have otherwise been validated only by
numerical simulations.  Unfortunately, however, since the analysis in
Refs.~\cite{schenkel,vanbaalen} relies on a comparison principle for
single parabolic equations~\cite{aronson,gmira} of the form
(\ref{eq:single}), it does not (as the authors indicate) appear to be
applicable when $D_A \neq D_B$ (which also leads to a moving reaction
front). Van Baalen et al.~\cite{vanbaalen} also remark that their
analysis is not easily extended to certain intermediate reaction
orders ($1 < m=n \leq 3$). These difficulties are reflected in Koza's
recent studies of the general cases $D_A > D_B > 0$ ~\cite{koza1} and
$D_B = 0$ ~\cite{koza2}, in which several {\it ad hoc} (but
reasonable) approximations are made and transients are ignored.

In the present article, the long-time asymptotics of the
initial-boundary-value problem (\ref{eq:rhoBC})--(\ref{eq:rho}) are
studied. This special case of (\ref{eq:coup}) is more tractable
analytically than the general case because (\ref{eq:rhoB}) can be
integrated exactly in time, thereby reducing the coupled system to a
single integro-partial differential equation. This useful
simplification is presented in section~\ref{sec:prelim} where the
problem is recast in a dimensionless form. It is also noted that
similarity solutions are expected to exist because there is no natural
length or time scale in the
problem~\cite{barenblatt96,barenblatt87,dresner}. Although it may be
possible to prove that the system actually approaches such a
self-similar solution starting from the prescribed initial conditions,
we instead pursue the more modest goal of proving that if an
asymptotic similarity solution exists, it must have a certain unique
form, {\it i.e.}  we explore the consequences of the
``quasi-stationary approximation''~\cite{cornell,koza1,koza2}.  In
section~\ref{sec:similarity}, the similarity solution is
systematically derived, and it is shown that a \lq\lq diffusion
layer'' (where the reaction term is dominated by the diffusion term)
with different scaling properties than the \lq\lq reaction front''
(where the reaction and diffusion terms balance) must exist to satisfy
the boundary conditions.  In section~\ref{sec:transients}, the
transient decay of the reaction rate in the diffusion layer is
analyzed, thereby proving {\it a posteriori} that the reaction term
can indeed be neglected in the dominant balance.  In
section~\ref{sec:uniform}, a uniformly-valid, asymptotic approximation
is constructed by matching the self-similar forms in the two different
regions. Finally, in section~\ref{sec:disc} some general physical
conclusions are drawn from the analysis,
and in the Epilogue certain similarities are discussed between this
work and the literature on combustion waves.
(Note that section \ref{sec:transients} is more technical and may be
skipped in a first reading.)

\section{Preliminaries}
\label{sec:prelim}

\subsection{Dimensionless Formulation}
\label{sec:dim}

With the definitions,
\begin{subeqnarray}
t \equiv m^\prime k (\rho_A^o)^{m-1}(\rho_B^o)^n T, \ & \ & \ x \equiv
X \sqrt{m^\prime k (\rho_A^o)^{m-1}(\rho_B^o)^n/D_A} \slabel{eq:units}
\\ a(x,t) \equiv \rho_A(X,T)/\rho_A^o, \ & \ & \ b(x,t) \equiv
\rho_B(X,T)/\rho_B^o ,
\end{subeqnarray}
the initial-boundary-value problem (\ref{eq:rhoBC})--(\ref{eq:rho})
may be expressed in a dimensionless form
\begin{subeqnarray}
\frac{\partial a}{\partial t} & = & \frac{\partial^2 a}{\partial x^2} -
a^mb^n \slabel{eq:a} \\
\frac{\partial b}{\partial t} & = & - q a^mb^n . \slabel{eq:b} \\
a(\infty,t) = 1, \ b(\infty,t) = 0, & \ & a(-\infty,t) = 0,
\ b(-\infty,t) = 1 , \slabel{eq:bc} \\
a(x,0) = H(x), & \ & b(x,0) = H(-x) \slabel{eq:ic}
\label{eq:eqs}
\end{subeqnarray}
which involves only one dimensionless parameter:
\begin{equation}
q \equiv \frac{n^\prime\rho_A^o}{m^\prime\rho_B^o} .
\end{equation}
Note that the dimensionless problem (\ref{eq:eqs}) depends only upon
the initial concentrations $\rho_A^o$ and $rho_B^o$ and the
stochiometric coefficients $m^\prime$ and $n^\prime$ through the
parameter $q$; the reaction rate $k$ are the diffusion constant $D_A$
simply set the natural scales for length and time. From
(\ref{eq:units}), we see that the limit of ``fast reactions'' $k
\rightarrow \infty$ (with $X$ and $T$ fixed) corrseponds to the limit
of long (dimensionless) times $t\rightarrow\infty$ at the diffusive
scale $x \propto \sqrt{t}$. (See Ref.~\cite{peletier96} for another
discussion of this correspondence of limits.)

\subsection{The Governing Integro-Partial Differential Equation}

The statement of the problem (\ref{eq:eqs}) will be used in deriving
the asymptotic similarity solution below, but for the transient
analysis described in section ~\ref{sec:transients} it will be
convenient to first integrate (\ref{eq:b}) exactly in time.  Note that
(\ref{eq:b}) and (\ref{eq:ic}) imply that $b(x,t) = 0$ for $x>0$ at
all times $t \geq 0$, which reflects the fact that species B cannot
diffuse out of its initial region. For $x < 0$, we integrate
(\ref{eq:b}) using the initial condition (\ref{eq:ic}) to express
$b(x,t)$ as
\begin{equation}
b(x,t) = \left\{ \begin{array} {ll}
e^{-q \phi_m(x,t)} & \ \mbox{if} \ n=1 \\
\left[ 1 + q(n-1)\phi_m(x,t) \right]^{-1/(n-1)} & \ \mbox{if} \
n \neq 1
\end{array}
\right. , \ \ \ x < 0 \label{eq:bint}
\end{equation}
which involves the time-integral of $a(x,t)^m$:
\begin{equation}
\phi_m(x,t) \equiv \int_0^t a(x,\tau)^m d\tau . \label{eq:phidef}
\end{equation}
(Note that a partial differential equation satisfied by $\phi_1(x,t)$
is given in Ref.~\cite{peletier96}.)
Substituting for $b(x,t)$ in (\ref{eq:a}), we obtain a single,
nonlinear integro-partial differential equation for $a(x,t)$, either
\begin{equation}
\frac{\partial a(x,t)}{\partial t} = \frac{\partial^2
a(x,t)}{\partial x^2} - H(x) a(x,t)^m e^{-q\int_0^t a(x,\tau)^m
d\tau}   \label{eq:ide1}
\end{equation}
if $n=1$ or
\begin{equation}
\frac{\partial a(x,t)}{\partial t} = \frac{\partial^2
a(x,t)}{\partial x^2} - \frac{H(x) a(x,t)^m}{\left[ 1 + q(n-1)\int_0^t
a(x,\tau)^m d\tau \right]^{n/(n-1)}}   \label{eq:ide2}
\end{equation}
if $n \neq 1$. 

Although these equations involve only one unknown function $a(x,t)$,
they are somewhat unwieldy, so we will first seek long-time ($t
\rightarrow \infty$) asymptotic solutions to the coupled system
(\ref{eq:eqs}) in section~\ref{sec:similarity}.  The time-dependent
properties of (\ref{eq:ide1}) and (\ref{eq:ide2}) will be studied in
section~\ref{sec:transients}. Before proceeding, however, we digress
to show that physically meaningful solutions exist only if $n \geq
1$. Later in the analysis we will see that $m \geq 1$ is required as
well.

\subsection{A Reaction Front Does Not Exist if $n < 1$}
\label{sec:exist_n}

Consider any point $x_o < 0$. Since species A diffuses to $x_o$ from a
reservoir of constant concentration ($a(\infty,t) = 1$) while species
B is removed by reactions without ever being replenished ($\partial
b(x_o,t)/\partial t < 0$ for all $t > 0$), it is clear that after long
times $a(x_o,t)$ must eventually differ from zero. Therefore, there
exists some $a_\ast(x_o) > 0$ and $t_o > 0$ such that $a(x_o,t) >
a_\ast(x_o)$ for all $t > t_o$. This implies $\phi_m(x_o,t) >
a_\ast(x_o)^m \cdot(t-t_o)$ from (\ref{eq:phidef}) and thus $b(x_o,t)
\rightarrow 0$ from (\ref{eq:bint}) since $q>0$, but a singularity
arises if $n < 1$: The concentration of static reactant $b(x_0,t)$
vanishes at some finite time $t_1$ given by $q(1-n)\phi_m(x_o,t_1)=1$
(which exists because $\phi(x_o,t)$ is continuous, $\phi_m(x_o,0)=0$
and $\phi_m(x_o,\infty)=\infty$). For $t > t_1$, Eq.~(\ref{eq:bint})
predicts imaginary, negative, or diverging solutions for $n < 1$, none
of which are physically meaningful. Therefore, when $n<1$, the
solutions to the model equations break down physically in a finite
time, and in that sense there does not exist a stable, moving reaction
front.  Nevertheless, it should be noted that Hilhorst et
al.~\cite{peletier96,peletier99} have shown that well-defined
solutions with free boundaries (at the diffusive scale $x \propto
\sqrt{t}$) can exist when $0 < n < 1$ .

\section{Derivation of the Asymptotic Similarity Solution}
\label{sec:similarity}

\subsection{Scaling of the Reaction Front}
\label{sec:Ac}

\begin{figure}[htb]
\begin{center}
\mbox{
\psfig{file=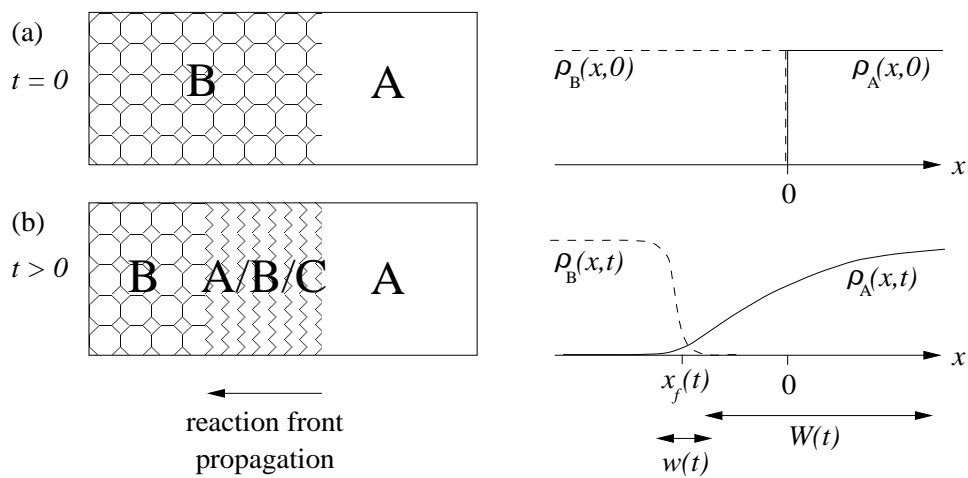,width=5in}
}
\begin{minipage}[h]{6in}
\caption{ Schematic diagrams (on the left) and concentration sketches (on
the right) showing (a) the diffusing reactant A initially separated
from the static reactant B and (b) the formation of a reaction front
propagating into the region rich in species B at some later time
leaving in its wake a mixture of species A and the reaction product
C. Various quantities discussed in section~\ref{sec:similarity} are
also indicated, {\it e.g.} the position of the reaction front $x_f(t)$
and the widths of the diffusion layer $W(t)$ and reaction front
$w(t)$.
\label{fig:cartoon}
}
\end{minipage}
\end{center}
\end{figure}

The initial-boundary-value problem (\ref{eq:eqs}) possesses no natural
length or time scale, {\it i.e.} it is invariant under power-law
\lq\lq stretching transformations''~\cite{dresner}, and consequently
in the limit $t \rightarrow \infty$ the system is expected to approach
an asymptotic similarity solution in which distance and time are
coupled by power-law scalings~\cite{barenblatt96,barenblatt87}. Since
reactant A diffuses while reactant B does not, the (presumably unique)
point of maximal reaction rate $r(a,b) = a^mb^n$ moves in the $-x$
direction toward the reservoir of reactant B. Therefore, an asymptotic
similarity solution, if one exists, must involve a moving frame of
reference centered on some point $x_f(t)$ identifying the position of
the reaction front at or near the point of maximal reaction rate (with
$dx_f/dt < 0$). Let $x_f(t) = -2 \nu t^\sigma$, where $\nu(q) > 0$ is
a constant (akin to the \lq\lq speed'' of the front) to be determined
self-consistently during the analysis, and consider an arbitrary
coordinate stretching transformation in the moving reference frame,
\begin{equation}
\eta \equiv \frac{x + 2\nu t^\sigma}{t^\alpha} .
\end{equation}
where $w(t) = t^\alpha$ is the width of the reaction front indicated
in Fig.~\ref{fig:cartoon}. (The factor of two is included only for
algebraic convenience.)

In the neighborhood of $x_f(t)$, we
also allow the magnitude of $a(x,t)$ to vary with a power-law scaling,
\begin{equation}
\tilde{\Ac}(\eta,t) \equiv t^{\gamma} a(x,t) . \label{eq:Acdef}
\end{equation}
If $\gamma \neq 0$, then another similarity solution far away from the
reaction front (in the \lq\lq diffusion layer'' shown in
Fig.~\ref{fig:cartoon} and defined below) will be needed to satisfy
the boundary condition $a(\infty,t)=1$. This possibility that {\it two
regions with different asymptotically self-similar dynamics for
$a(x,t)$} could arise is suggested by the fact that there are two
driving terms, representing diffusion and reaction, on the right-hand
side of (\ref{eq:a}) with different behaviors under stretching
transformations. On the other hand, there is only the reaction term on
the right-hand side of (\ref{eq:b}), so {\it $b(x,t)$ can exhibit
only one type of asymptotic scale invariance}. This is the main
mathematical consequence of the physical fact that reactant B does not
diffuse. Since $b(-\infty,1)=1$, we consider the transformation
\begin{equation}
\tilde{\Bc}(\eta,t) \equiv b(x,t) .
\end{equation}
Note that the reaction term $r(a,b) = a^mb^n$ has the scaling, $r
= t^{-\beta}\tilde{\Ac}^m\tilde{\Bc}^n$, where $\beta = m\gamma$ in
the notation of G\'alfi and R\'acz~\cite{galfi}. 

These transformations leave the governing equations in the form:
\begin{subeqnarray}
t^{(m-1)\gamma} \frac{\partial \tilde{\Ac}}{\partial t}
- t^{(m-1)\gamma-1} \left(\gamma\tilde{\Ac} + \alpha\eta\frac{\partial
\tilde{\Ac}}{\partial \eta}\right) & & \nonumber \\
 + t^{(m-1)\gamma-1-\alpha+\sigma} 2\sigma\nu
\frac{\partial \tilde{\Ac}}{\partial \eta}
& = & t^{(m-1)\gamma-2\alpha}
\frac{\partial^2 \tilde{\Ac}}{\partial \eta^2} - \tilde{\Ac}^m \tilde{\Bc}^n ,
        \slabel{eq:Aclong} \\
t^{m\gamma} \frac{\partial \tilde{\Bc}}{\partial t}
- t^{m\gamma-1} \alpha \eta \frac{\partial \tilde{\Bc}}{\partial \eta}
+ t^{m\gamma-1-\alpha+\sigma} 2\sigma\nu \frac{\partial \tilde{\Bc}}{\partial
\eta}  & = &  - q \tilde{\Ac}^m \tilde{\Bc}^n  . \slabel{eq:Bclong}
\label{eq:AcBclong}
\end{subeqnarray}
We now look for asymptotically invariant solutions
\begin{subeqnarray}
& & \tilde{\Ac}(\eta,t) \rightarrow \Ac(\eta), \ \ \ \frac{\partial
\tilde{\Ac}}{\partial \eta}(\eta,t)
\rightarrow \Ac^\prime(\eta), \ \ \ \frac{\partial^2
\tilde{\Ac}}{\partial \eta^2}(\eta,t) \rightarrow
\Ac^{\prime\prime}(\eta)  \slabel{Ac_converge}  \\
& & \tilde{\Bc}(\eta,t) \rightarrow \Bc(\eta), \ \ \ \frac{\partial
\tilde{\Bc}}{\partial \eta}(\eta,t)
\rightarrow \Bc^\prime(\eta)
\label{eq:AcBc_converge}
\end{subeqnarray}
as $t \rightarrow \infty$ with $|\eta| < \infty$ fixed.  For
consistency with our definition of the reaction front, we require that
there is in each equation a dominant balance between the reaction term
$\Ac^m \Bc^n$ and at least one other non-vanishing term. In order for
time invariance to be attained, we assume that time-dependent terms in
the transformed coordinates are negligible compared to the reaction
term, {\it i.e.}
\begin{equation}
\lim_{t\rightarrow\infty}t^{(m-1)\gamma} \frac{\partial \tilde{\Ac}}{\partial
t}
= \lim_{t\rightarrow\infty}t^{m\gamma} \frac{\partial \tilde{\Bc}}{\partial t}
= 0 , \label{eq:rfstat}
\end{equation}
which is a precise statement of the assumption of \lq\lq
quasi-stationarity''~\cite{koza2}.

A dominant balance with the reaction term in (\ref{eq:Aclong})
implies that at least one of the following three cases must be true:
\begin{description}
\item[Case A1:] $\ \ (m-1)\gamma-2\alpha = 0, \ \
(m-1)\gamma-1-\alpha+\sigma\leq 0, \ \ (m-1)\gamma-1\leq 0,$
\item[Case A2:] $\ \ (m-1)\gamma-2\alpha \leq 0, \ \
(m-1)\gamma-1-\alpha+\sigma= 0, \ \ (m-1)\gamma-1\leq 0,$
\item[Case A3:] $\ \ (m-1)\gamma-2\alpha\leq 0, \ \
(m-1)\gamma-1-\alpha+\sigma\leq 0, \ \ (m-1)\gamma-1= 0.$
\end{description}
Likewise a dominant balance in (\ref{eq:Bclong}) requires that one
of the following two cases must hold:
\begin{description}
\item[Case B1:] $\ \ m\gamma-1 = 0, \ \ m\gamma-1-\alpha+\sigma\leq 0,$
\item[Case B2:] $\ \ m\gamma-1 \leq 0, \ \ m\gamma-1-\alpha+\sigma = 0.$
\end{description}
There are only two combinations of these cases that are logically
consistent:
\begin{description}
\item[Traveling Wave Case:] (A2,\ B2) $\ \ \alpha \geq 0,\ \
\gamma = 0, \ \ \sigma = 1 + \alpha$, and
\item[Diffusing Front Case:] (A1,\ B2) $\ \ \alpha = (m-1)\gamma/2, \
\ \sigma = 1 - (m+1)\gamma/2, \ \ 0 < \gamma \leq 1/m$.
\end{description}
In the first case, we have $\sigma \geq 1 $, which implies that the
reaction front advances at least linearly, {\it e.g.} as a traveling
wave $x_f \sim t$, but in the second case, the front advances
sublinearly, {\it e.g.} as a diffusing front $x_f \sim t^{1/2}$. In
both cases, note that the reaction order $n$ of the static species B
plays no role in the scaling behavior. The same conclusion is also
true of the reaction order $m$ of the diffusing species A in the
Traveling Wave Case, but $m$ does affect the scaling exponents in the
Diffusing Front Case.

Consider the possibility $\gamma=0$, which is only
consistent with the Traveling Wave Case. In this case, a single
asymptotic scale invariance is attained everywhere, and the equations
for $\Ac(\eta)$ and $\Bc(\eta)$ are
\begin{subeqnarray}
2\sigma\nu \Ac^\prime & = & \delta_{\alpha,0} \Ac^{\prime\prime} -
        \Ac^m\Bc^n \\
2\sigma\nu \Bc^\prime & = & -q \Ac^m\Bc^n 
\end{subeqnarray}
where $\delta_{x,y}$ is the Kronecker delta.  By combining these
equations and integrating once using the boundary conditions behind
the front, {\it i.e.} $\Ac(\infty) = 1$, $\Ac^\prime(\infty)=0$ and
$\Bc(\infty)=0$, we obtain
\begin{equation}
2\sigma\nu(\Bc + q) = q (2\sigma\nu \Ac - \delta_{\alpha,0}
        \Ac^{\prime}) .
\end{equation}
Applying the boundary conditions ahead of the front,
$\Ac(-\infty)=\Ac^{\prime}(-\infty)=0$ and $\Bc(-\infty)=1$, to this
equation then implies $\sigma\nu(1+q) = 0$, which is a contradiction
since $\sigma > 0$ and $\nu > 0$ are needed for the reaction front to
move at all (and $q > 0$).

In this way, we are forced to consider at least {\it two regions with
different scale invariance} if there is to be any hope of an
asymptotic similarity solution. Since the second type of scale
invariance is associated with the dominance of the diffusion term
versus the reaction term in (\ref{eq:a}), it must occur only on
the back ($+x$) side of the reaction front due to the reservoir of
reactant A at infinity, $a(\infty,t) = 1$ (see
Fig.~\ref{fig:cartoon}). To describe the scale invariance of the
diffusion layer, we postulate another power law $W(t) = t^\delta$ for
the asymptotic width of the diffusion layer.

\subsection{Scaling of the Diffusion Layer}

Since $\delta \neq \alpha$, there are two possibilities, each
involving a singular perturbation $w/W = t^{\alpha-\delta}$:
\begin{description}
\item[Infinitely Thin Reaction Front Case:] \ $\delta > \alpha$, \ $w =
o(W)$,
\item[Infinitely Thin Diffusion Layer Case:] \ $\delta < \alpha$, \ $W =
o(w)$.
\end{description}
Since chemical reactions are typically much faster than diffusion, the
former case seems more reasonable on physical grounds, but we do not
rule out the latter case {\it a priori}. In the Infinitely Thin
Reaction Front Case, the reaction front is defined by $x-x_f =
O(w)$ and the diffusion layer by $W =
O(x-x_f)$, $x>x_f$, whereas in the Infinitely Thin Diffusion
Layer Case, the reaction front is defined by $w =
O(x-x_f)$, $x<x_f$ and the diffusion layer by $x-x_f =
O(W)$. In both cases, we view the reaction front as
representing the \lq\lq inner problem'' (with similarity variable
$|\eta|<\infty$) and the diffusion layer as representing the \lq\lq
outer problem'' (with similarity variable $\zeta > 0$ defined
below). The two regions are connected by asymptotic matching of the
limits $\eta\rightarrow \infty$ and $\zeta \rightarrow 0^+$ (described
in the next section) ~\cite{hinch,bender}.

In order to treat the outer problem, we transform the original
equations using a new reduced coordinate with power-law scalings,
\begin{equation}
\zeta \equiv \frac{x + 2\nu t^\sigma}{2 t^\delta} , \ \ \ \ 
 \tilde{A}(\zeta,t) \equiv a(x,t), \ \ \ \
\tilde{B}(\zeta,t) \equiv b(x,t) .
\label{eq:zeta}
\end{equation}
(Another factor of two is included in $\eta= 2t^{\delta-\alpha}\zeta$
for algebraic convenience. Note the use of $A$ and $B$ for the
diffusion layer versus $\Ac$ and $\Bc$ for the reaction front.)  Under
this transformation, the equations take the form,
\begin{subeqnarray}
t^{2\delta}\frac{\partial \tilde{A}}{\partial t} - t^{2\delta-1} \delta\zeta
\frac{\partial \tilde{A}}{\partial \zeta} +
t^{\delta+\sigma-1} \sigma\nu \frac{\partial \tilde{A}}{\partial \zeta} & = &
\frac{1}{4}\frac{\partial^2 \tilde{A}}{\partial \zeta^2} 
- t^{2\delta} \tilde{A}^m \tilde{B}^n ,
\slabel{eq:Along} \\
t^{2\delta}\frac{\partial \tilde{B}}{\partial t}
- t^{2\delta-1} \delta\zeta \frac{\partial \tilde{B}}{\partial \zeta} +
t^{\delta+\sigma-1} \sigma\nu \frac{\partial \tilde{B}}{\partial \zeta} & = &
- t^{2\delta} q \tilde{A}^m \tilde{B}^n .
\slabel{eq:Blong}
\label{eq:ABlong}
\end{subeqnarray}

Seeking an asymptotic similarity solution, we assume that invariance
is achieved in the transformed equations
\begin{subeqnarray}
& & \tilde{A}(\zeta,t) \rightarrow A(\zeta), \ \ \ \frac{\partial
\tilde{A}}{\partial \zeta}(\zeta,t)
\rightarrow A^\prime(\zeta), \ \ \ \frac{\partial^2
\tilde{A}}{\partial \zeta^2}(\zeta,t) \rightarrow
A^{\prime\prime}(\zeta)  \slabel{eq:A_converge}   \\
& & \tilde{B}(\zeta,t) \rightarrow B(\zeta), \ \ \ \frac{\partial
\tilde{B}}{\partial \zeta}(\zeta,t)
\rightarrow B^\prime(\zeta) \slabel{eq:B_converge}
\label{eq:AB_converge}
\end{subeqnarray}
assuming time-variations in (\ref{eq:A_converge}) are small relative
to the diffusion term
\begin{equation}
\lim_{t \rightarrow\infty} t^{2\delta}\frac{\partial
\tilde{A}}{\partial t}(\zeta,t)  = 0,\ \
\mbox{for}\ \zeta>0. \label{eq:dlstat}
\end{equation}
In order to obtain a different scaling from the reaction front, the
reaction term must also not enter into the dominant balance
\begin{equation}
\label{eq:Rzero}
\lim_{t\rightarrow\infty} t^{2\delta} \tilde{A}(\zeta,t)^m
\tilde{B}(\zeta,t)^n = 0  
\ \ \mbox{for $\zeta > 0$,}
\end{equation}
a condition that we will check {\it a posteriori} for consistency in
section~\ref{sec:transients}. Note that this limit vanishes trivially
for $\zeta > \nu$ since we have already noted that $b(x,t)=0$ for all
$x>0$. From (\ref{eq:Blong}), this condition would imply $\partial
\tilde{B}/\partial
\zeta = 0$, which together with the boundary condition $\tilde{B}(\infty) = 0$
would imply $\tilde{B}(\zeta) = 0$.  With the reaction term gone, one
of the terms on the left-hand side of (\ref{eq:Along}) must balance
the $\partial^2 \tilde{A}/\partial \zeta^2$ term on the right side; if
not, we would have $\partial^2 \tilde{A}/\partial \zeta^2=0$, which
cannot satisfy all of the boundary conditions. There are only two
possible dominant balances:
\begin{description}
\item[Case D1:] $\delta+\sigma-1 =0 \ \ \mbox{and} \ \
2\delta-1 \leq 0$,
\item[Case D2:] $\delta+\sigma-1 \leq 0 \ \ \mbox{and} \ \
2\delta-1 = 0$.
\end{description}
The former case implies $\sigma > 1/2$ and hence contains the
Traveling Wave Case (and not the Diffusing Front Case). With the
scaling relations of case D1, Eq.~(\ref{eq:Along}) has the asymptotic
form
\begin{equation}
\sigma\nu A^\prime = \frac{1}{4} A^{\prime\prime}.
\end{equation}
The solutions to this equation exhibit exponential growth as $\zeta
\rightarrow \infty$, which is incompatible
with the boundary condition $A(\infty) = 1$.  Therefore, we conclude
that Case D1, and hence also the Traveling Wave Case, is not
consistent with the boundary conditions. At this point, we are left
with Case D2 together with the Diffusing Front Case (A1,B2), which
imply $W \sim \sqrt{t}$, thus justifying the term \lq\lq diffusion
layer" for the region $\zeta > 0$.

\subsection{Asymptotic Matching of the Reaction Front and Diffusion
Layer}

One more condition is needed to uniquely determine the scaling
exponents, and it comes from asymptotic matching: The \lq\lq outer
limit'' $\eta \rightarrow \infty$ of the inner approximation must
match with the \lq\lq inner limit'' $\zeta \rightarrow 0^+$ of the
outer approximation (because both are asymptotic representations of
the same function). Unlike the more familiar case of boundary layers
of ordinary differential equations~\cite{hinch,bender}, however, our
system of  partial differential equations will require extra care
for matching because the limit $t \rightarrow \infty$ (with either
$\zeta$ or $\eta$ fixed) must be taken before the inner and
outer limits.

Since $B(\zeta) = 0$ for all $\zeta > 0$, the only matching condition
for $b(x,t)$ is trivial, $\Bc(\infty) = 0$, but the matching
conditions for $a(x,t)$ are more subtle.  Since $\gamma > 0$, the
concentration of species $A$ approaches $0$ in the reaction front:
$a(x,t) = O(t^{-\gamma})$ as $t
\rightarrow \infty$ with $|\eta|<\infty$ fixed. Therefore, a boundary
condition on the outer problem is $A(\zeta) = 0$, but unfortunately
this does not provide a boundary condition on the inner problem.
Instead, we must consider matching at the next (linear) order of
Taylor expansion in the intermediate region:
\begin{equation}
\frac{\partial a}{\partial x}  =  \left\{ \begin{array}{ll}
\frac{\partial \tilde{A}}{\partial \zeta} \frac{\partial
\zeta}{\partial x} \sim 
\frac{A^\prime(\zeta)}{2t^\delta} &  \ \  \mbox{as} \ t \rightarrow
\infty \ \mbox{with} \ 0 < \zeta < \infty \ \mbox{fixed}  \\
\frac{1}{t^\gamma} \frac{\partial \tilde{\Ac}}{\partial \eta}
\frac{\partial \eta}{\partial x} \sim
\frac{\Ac^\prime(\eta)}{t^{\alpha+\gamma}} & \ \  \mbox{as} \ t \rightarrow
\infty \ \mbox{with} \ |\eta| < \infty \ \mbox{fixed}  
\end{array} \right. .
\end{equation}
Now requiring that the two intermediate limits match yields the final
scaling relation:
\begin{equation}
\alpha + \gamma = \delta \label{eq:agd}
\end{equation}
as well as the missing boundary condition on the inner problem:
\begin{equation}
\Ac^{\prime}(\infty) = A^\prime(0)/2 .
\label{eq:fluxmatch}
\end{equation}
(Note that $A(\zeta)$ is already fully determined by matching at
zeroth order.) This scaling relation (\ref{eq:agd}) can be understood
physically as expressing conservation of mass between the diffusion
layer and reaction front \cite{jiang}. Similarly, the matching
condition (\ref{eq:fluxmatch}) simply means that the diffusive flux
entering the reaction front equals the flux leaving the diffusion
layer.

By examining all possible similarity solutions with power-law
couplings of distance and time, we finally arrive at a {\it unique}
set of scaling exponents from cases A1, B2 and D2 and
(\ref{eq:agd}):
\begin{equation}
\alpha = \frac{m-1}{2(m+1)},\ \ \beta = \frac{m}{m+1}, \ \
\gamma = \frac{1}{m+1}, \ \ \sigma = \delta = \frac{1}{2} .
\label{eq:exponents}
\end{equation}
Therefore, after long times the reaction front itself \lq\lq diffuses"
according to $x_f(t) = -2\nu\sqrt{t}$, where $\nu(q)^2$ is now
interpreted as an effective diffusion constant for the front.
Although for $m = 1$ the reaction zone settles down to a constant
width ($\alpha = 0$), for $m>1$ the front width grows in time ($\alpha
> 0$). In all cases the reaction front is \lq\lq infinitely thin"
compared to the diffusion layer ($\alpha < \delta$). Note that as $m$
increases, $\gamma$ tends to zero, meaning that the concentration in
the reaction front does not decrease as quickly for higher-order
reactions as it does for first-order reactions.  The exponents
$\alpha=0$ and $\gamma = 1/2$ for the case $m=1$ were first obtained 
by Jiang and Ebner~\cite{jiang} based on physical arguments supported
by Monte Carlo simulations and later discussed in an analytical
context by Koza~\cite{koza2}, but to our knowledge prior to this work
neither have the general expressions for $m \neq 1$ been given nor
have the scaling exponents been proven to be unique.  With the scaling
exponents and matching boundary conditions now determined, we proceed
to solve the inner and outer boundary-value problems in the following
sections.

\subsection{Concentration Profiles in the Diffusion Layer}
\label{sec:diff}

\begin{figure}[htb]
\begin{center}
\mbox{
\psfig{file=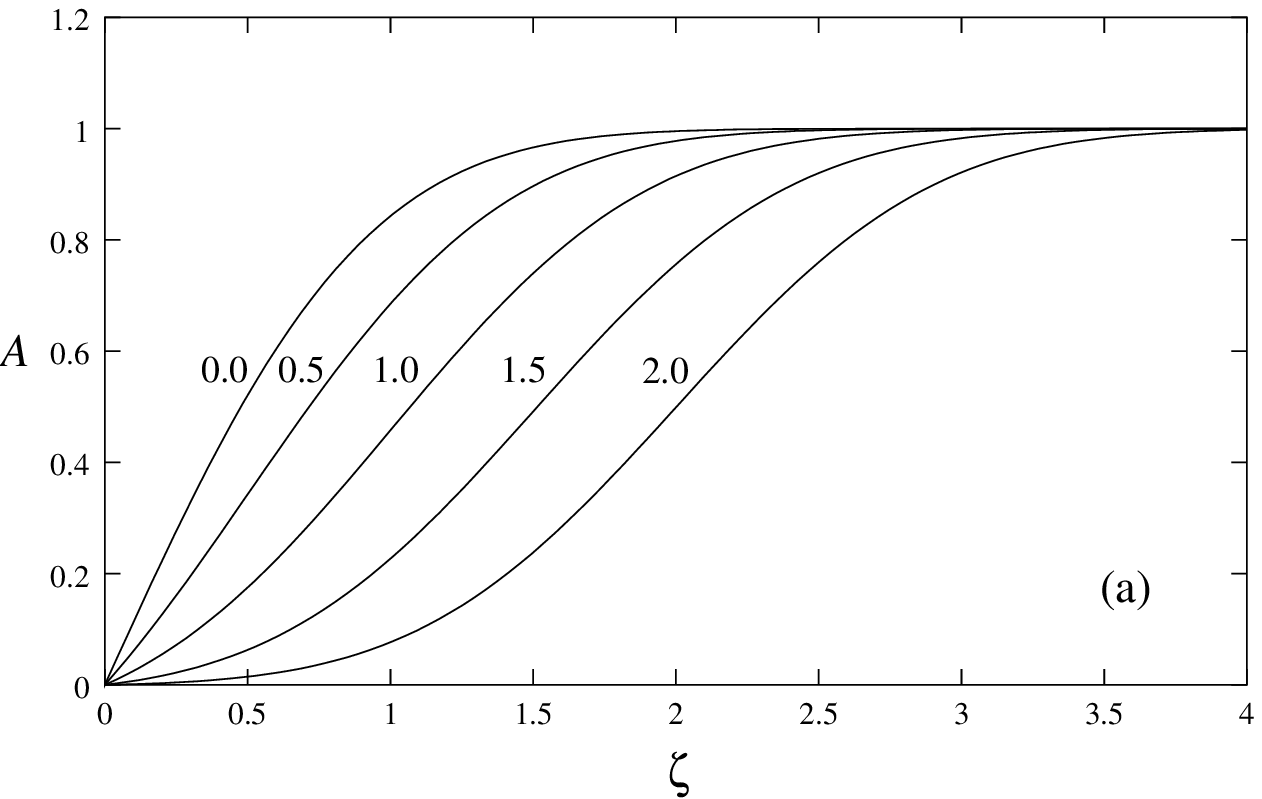,height=3in}
}
\mbox{
\psfig{file=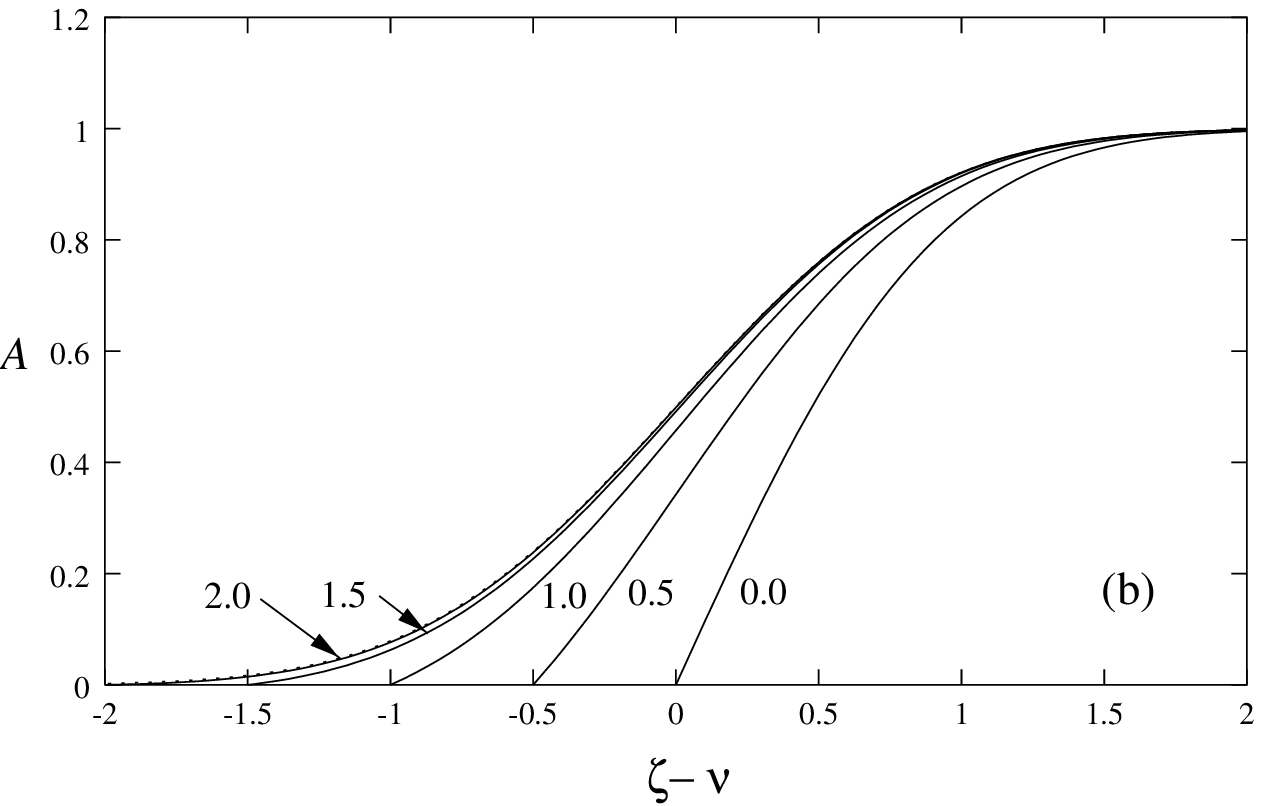,height=3in}
}
\begin{minipage}[h]{5.5in}
\caption{The asymptotic similarity function in the diffusion layer,
$a(x,t) \sim A(\zeta)$ where $\zeta = \nu + x/2\protect\sqrt{t}$ shown
for $\nu = 0.0, 0.5, 1.0, 1.5, 2.0$ versus $\zeta$ in (a) and versus $\zeta
- \nu$ in (b). The limiting shape $\erf(x/2\protect\sqrt{t})$
corresponds to $\nu = 0$. The other limiting shape $[1 +
\erf(x/2\protect\sqrt{t})]/2$ as $\nu \rightarrow \infty$ is plotted
as the dashed line in (b), but it is almost indistinguishable from the
$\nu=2$ curve.  
\label{fig:A}
}
\end{minipage}
\end{center}
\end{figure}

In the diffusion layer from (\ref{eq:ABlong})--(\ref{eq:Rzero}) we
have
\begin{equation}
-2(\zeta - \nu)A^{\prime} = A^{\prime\prime}, \ \ \
 A(0)=0, \ A(\infty)=1 . \label{eq:Aeq}
\end{equation}
The exact solution to this boundary value problem can be expressed in
terms of error functions~\cite{abra}
\begin{equation}
A(\zeta) = \frac{\erf(\zeta-\nu) + \erf(\nu)}{1 + \erf(\nu)} 
\label{eq:diffA}
\end{equation}
and is depicted in Fig.~\ref{fig:A}.  Note that the dimensionless flux
entering the reaction front
\begin{equation}
\Ac_1(\nu) \equiv \frac{A^\prime(0)}{2} =
\frac{e^{-\nu^2}}{\sqrt{\pi}(1 + \erf(\nu))}  \label{eq:A1}
\end{equation}
is needed for asymptotic matching in (\ref{eq:fluxmatch}).

The effect of varying $q = n^\prime \rho_A^o/(m^\prime\rho_B^o)$ is
easily understood in terms of the mathematical model.  As $q$ is
decreased, the reaction front slows down since reactions in the front
region remove species A much faster than diffusion can replenish
it. In the limit $q\rightarrow 0$, {\it i.e.} $\rho_A^o\rightarrow 0$,
the front comes to a complete stop, $\nu(0) = 0$. For very small, but
finite $q > 0$, the concentration of diffusing reactant approximately
obeys 
\begin{equation}
\frac{\partial a}{\partial t} = \frac{\partial^2 a}{\partial x^2}, \ \ 
\ a(x,0) = H(x), \ \ \ a(0,t) = 0,  \ \ a(\infty,t) = 1 , \ \ x \geq 0
\end{equation}
(at least for $t \ll \nu^{-2}$ since the reaction front is stationary
only for short times).  This classical diffusion problem has the exact
similarity solution
\begin{equation}
a(x,t) = \erf\left(\frac{x}{2\sqrt{t}}\right) ,  \label{eq:Anuzero}
\end{equation}
which is precisely the $\nu=0$ curve in Fig.~\ref{fig:A}. From
(\ref{eq:diffA}) note that even when the front has moved
significantly ($t \gg \nu^{-2}$) the concentration still has the same
shape, $A(\zeta) \approx \erf(\zeta)$, in the (very slowly) moving
reference frame, as long as $\nu(q) \ll 1$.

On the other hand, as $q$ is increased, the reaction term becomes
progressively less important compared to the diffusion term in
(\ref{eq:a}). In the limit $q \rightarrow \infty$, {\it i.e.}
$\rho_B^o \rightarrow 0$, we recover another classical diffusion
problem (after sufficiently long times $t \gg \nu^{-2}$)
\begin{equation}
\frac{\partial a}{\partial t} = \frac{\partial^2 a}{\partial x^2}, \ \ 
\ a(x,0) = H(x),\ \ \ a(-\infty,t) = 0, \ \ a(\infty,t) = 1 ,
\end{equation}
which has the exact similarity solution
\begin{equation}
a(x,t) = \frac{1}{2} \left[ 1 +
\erf\left(\frac{x}{2\sqrt{t}}\right)\right]
\label{eq:Anuinfty}
\end{equation}
Note that in the limit $q \rightarrow \infty$ the reaction front
instantly speeds off to $-\infty$ having consumed only a negligible
amount of reactant A, resulting in a pure diffusion problem for
$a(x,t)$. Indeed, we will see below that $\nu(\infty) = \infty$. It
remains, of course, to relate $\nu$ and $q$.

In Fig.~\ref{fig:A}(b), we see how the true asymptotic similarity
solution $A(\zeta)$ interpolates between the limiting forms
(\ref{eq:Anuzero}) and (\ref{eq:Anuinfty}) as $\nu$ goes from $0$ to
$\infty$, respectively. It turns out that for $\nu \geq 2$ the
asymptotic behavior of the original reaction-diffusion system is
almost indistinguishable from (\ref{eq:Anuinfty}) for $t \gg 1/4$.

\subsection{Diffusion Constant of the Reaction Front}

In the reaction front, we have a third-order system of nonlinear
ordinary differential equations
\begin{subeqnarray}
0 & = & \Ac^{\prime\prime} - \Ac^m \Bc^n \slabel{eq:Ac} \\
\nu \Bc^\prime & = & - q \Ac^m \Bc^n , \slabel{eq:Bc}
\label{eq:AcBc}
\end{subeqnarray}
with four boundary conditions
\begin{equation}
\Ac(-\infty) = 0,\ \, \Bc(-\infty) = 1, \ \ \Bc(\infty) = 0, \ \
\mbox{and} \ \
\Ac^\prime(\infty) = \Ac_1 ,
\end{equation}
where $\Ac_1(\nu)$ is known via (\ref{eq:A1}).  Although this
boundary-value problem appears to be overdetermined, the fourth
boundary condition is actually necessary to determine the unknown
diffusion constant of the reaction front $\nu(q)$. 

By comparing (\ref{eq:AcBclong}) and (\ref{eq:AcBc}), some physical
insight into the dynamics of the reaction front is gained. The
concentration of diffusing reactant A is determined by a local balance
of reactions and \lq\lq steady state" diffusion and the concentration
of static reactant B by a local balance of reactions and fictitious
advection due to the translating reference frame. The latter balance
reflects the special character of $D_B = 0$: Since reactant B cannot
diffuse to the front, instead the front must diffuse to it. This is no
longer true if $D_B>0$ (no matter how small), which explains why
different scaling exponents arise in that case~\cite{galfi,jiang}.
These physical properties are manifested in the mathematical model by
the fact that since it multiplies the highest derivative in the
equations $D_B > 0$ is a singular perturbation.

\begin{figure}[htb]
\begin{center}
\mbox{
\psfig{file=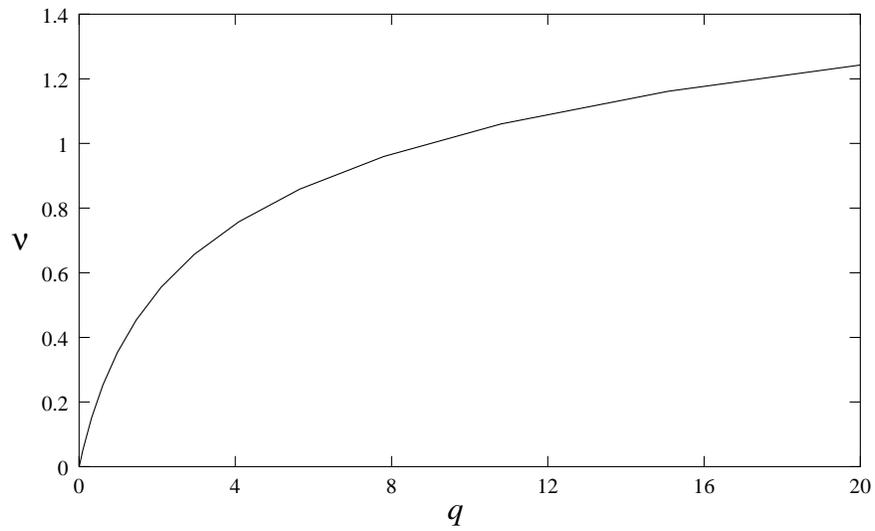,width=4.5in}
}
\begin{minipage}[h]{5.5in}
\caption{
The exact dependence of $\nu$, (the square root of) the dimensionless
diffusion constant of the reaction front, on the parameter $q =
n^\prime \rho_A^o/m^\prime\rho_B^o$ from Eq.~(\protect\ref{eq:qnu}). 
\label{fig:nuq}
}
\end{minipage}
\end{center}
\end{figure}

One integration of (\ref{eq:AcBc}) is easy to perform and fortunately
suffices to derive an exact expression for $\nu(q)$.  Substituting
(\ref{eq:Bc}) into (\ref{eq:Ac}), integrating and applying the
boundary conditions at $\eta = \infty$, we obtain
\begin{equation}
\nu \Bc = q (\Ac_1 - \Ac^\prime) .
\end{equation}
Likewise enforcing the boundary conditions at $\eta = -\infty$, we
find $\nu = q \Ac_1$. Substituting $\Ac_1$ from (\ref{eq:A1}), we have
\begin{equation}
\nu(q) = F^{-1}(q) ,
\label{eq:qnu}
\end{equation}
where
\begin{equation}
F(x) \equiv \sqrt{\pi} x e^{x^2}\left[ 1 + \erf(x)\right] ,
\end{equation}
which was first derived by Koza~\cite{koza2}. The function
$\nu(q)$ is plotted in Fig.~\ref{fig:nuq}.

The transcendental function $F(x)$ cannot be inverted analytically,
but limiting formulae can be derived. The Maclaurin series of $F(x)$
is
\begin{equation}
F(x) = \sqrt{\pi} x + 2 x^2 + \sqrt{\pi} x^3 + \frac{4}{3}x^4 +
\frac{\sqrt{\pi}}{2}x^5 + \frac{8}{15}x^6 + \frac{\sqrt{\pi}}{6}x^7 + \ldots ,
\end{equation}
which can be inverted term by term to generate the Maclaurin series of
$\nu(q)$, valid for small $q$,
\begin{equation}
\nu(q) = \frac{1}{\pi^{1/2}} q - \frac{2}{\pi^{3/2}}q^2 + \frac{8 -
\pi}{\pi^{5/2}}q^3 - \ldots .
\end{equation}
For large $q$, approximations such as,
\begin{equation}
\nu(q) \sim \sqrt{\log\left(\frac{q}{\sqrt{\pi}}\right) - \log 2
        - \frac{1}{2}\log\log\left(\frac{q}{\sqrt{\pi}}\right) } ,
\end{equation}
can be generated by iteration.

\subsection{Existence and Uniqueness of the Reaction-Front Scaling
Functions}
\label{sec:exist}

With the results of the previous section, the inner boundary-value
problem is reduced to a nonlinear, second-order equation for $\Ac(\eta)$:
\begin{equation}
\Ac^{\prime\prime} = \Ac^m(1 - \Ac^\prime/\Ac_1)^n,
\ \ \ \Ac(-\infty)=0, \ \ \Ac^\prime(\infty)=\Ac_1 .
 \label{eq:AcODE}
\end{equation}
Once this system is solved, $\Bc(\eta)$ is recovered from $\Bc(\eta) = 1 -
\Ac^\prime(\eta)/\Ac_1$. Note that (\ref{eq:AcODE}) is
invariant under translation $\eta \mapsto
\eta - \eta_o$, where the arbitrary constant $\eta_o$ sets the precise
location of the reaction front. Since $\eta_o$ depends on the exact
initial conditions, however, it cannot be determined by considering
only the long-time asymptotic limit as we have done here.

\begin{figure}[htb]
\begin{center}
\mbox{
\psfig{file=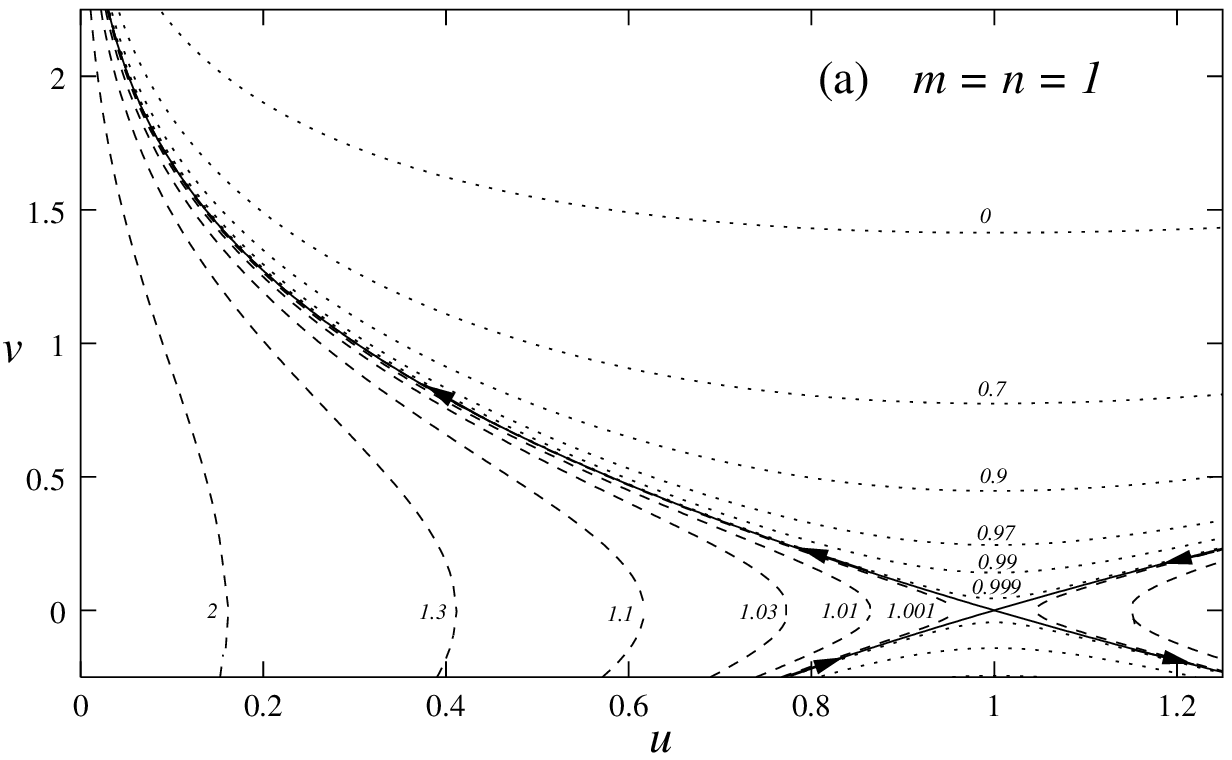,height=2.6in}
}
\mbox{
\psfig{file=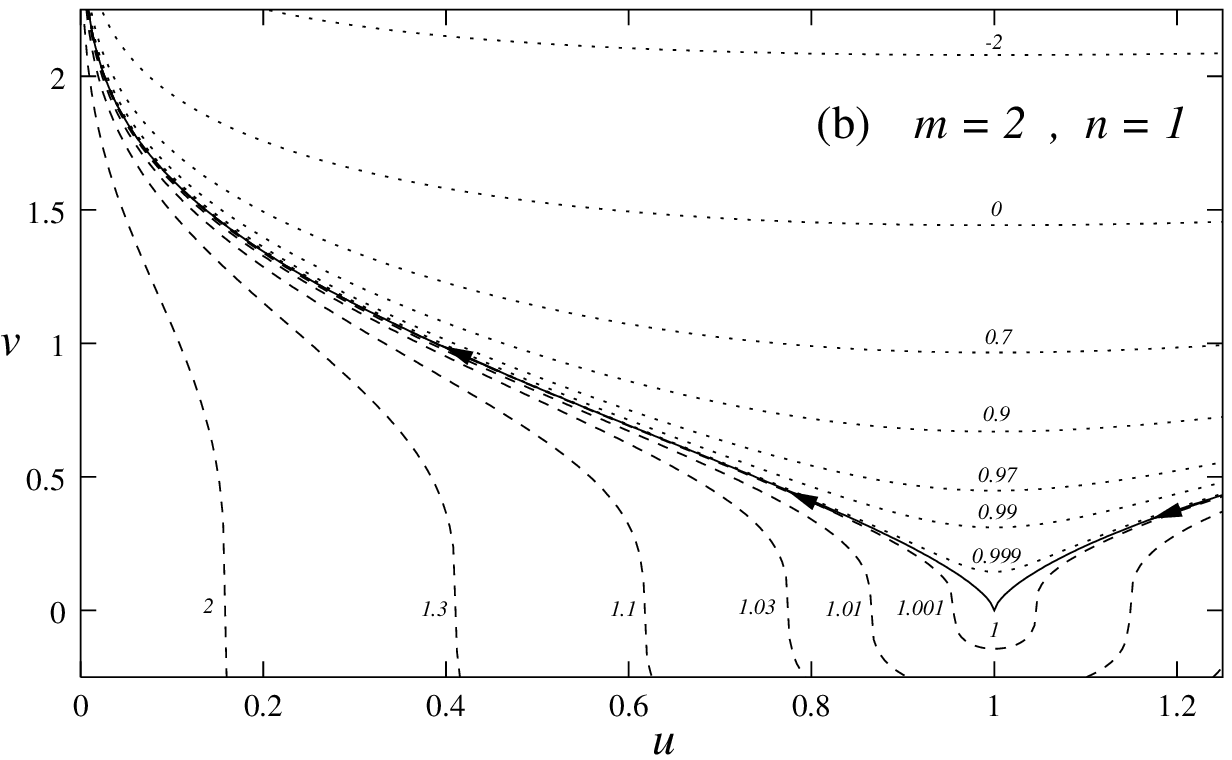,height=2.6in}
}
\mbox{
\psfig{file=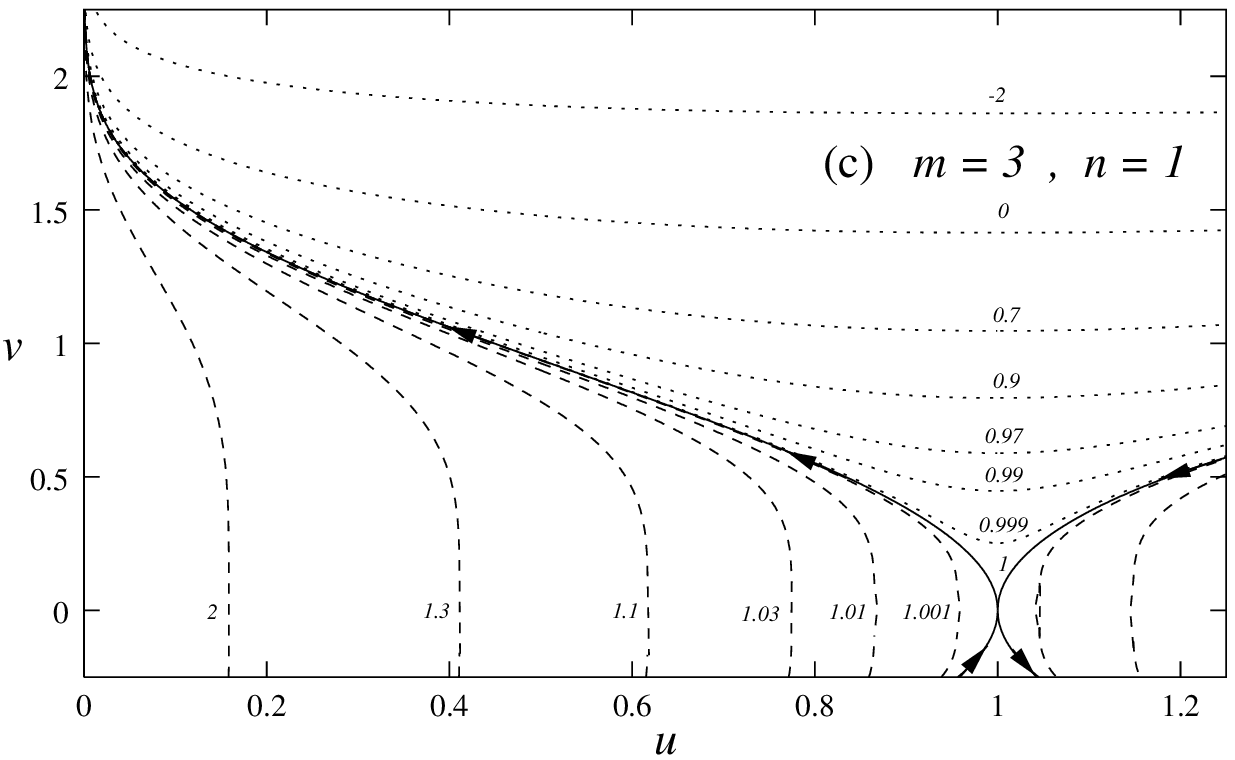,height=2.6in}
}
\begin{minipage}[h]{5.5in}
\caption{Phase-plane trajectories for the
inner (reaction-front) boundary-value problem from
Eq.~(\protect\ref{eq:Lie}) labeled by the constant $c_n$ for $m=n=1$
in (a), $m=2,n=1$ in (b) and $m=3,n=1$ in (c).  In each case, the
solid lines are separatrices emanating from the fixed point $(1,0)$ of
which the solution to the inner problem corresponds to the unique
curve connecting $(1,0)$ and $(0,\infty)$.  Arrows indicate the
direction of increasing $s$.
\label{fig:Lie}
}
\end{minipage}
\end{center}
\end{figure}

Since the second-order equation (\ref{eq:AcODE}) is autonomous ({\it
i.e.} $\eta$ does not appear), it is useful to consider the
\lq\lq Lie diagram''~\cite{dresner} or \lq\lq phase
plane''~\cite{bender,davis} of trajectories in the $(\Ac,\Bc)$ plane
parameterized by $\eta$, as shown in Fig.~\ref{fig:Lie}. By studying
properties of the phase plane, it is straightforward to prove the
existence and uniqueness of solutions if and only if $m,n
\geq 1$ and $\Ac_1 > 0$. With the change of variables
\begin{equation}
s \equiv \Ac_1^{(m-1)/(m+1)} \eta,
\ \ \ u(s) \equiv \Bc(\eta),  \ \ \ v(s) \equiv \Ac_1^{-2/(m+1)}
\Ac(\eta),  \label{eq:uvdef}
\end{equation}
we begin by transforming (\ref{eq:AcODE}) into a system of first-order
equations
\begin{subeqnarray}
u^\prime & = & - v^m u^n  \slabel{eq:uODE} \\
v^\prime & = & 1 - u \slabel{eq:vODE}
\label{eq:uv}
\end{subeqnarray}
with boundary conditions $v(-\infty)=0$ and $u(\infty) = 0$, or
equivalently $u(-\infty)=1$ and $v^\prime(\infty)=1$.

There is a unique fixed point at $(u,v) = (1,0)$ corresponding to the
region ahead of the reaction front which contains only the static
reactant B. This is the starting point ($s=-\infty$) of any
trajectories that satisfy the boundary condition $v(-\infty)=0$, so
our task is to identify and follow any unstable manifolds leaving
$(1,0)$ to see if they satisfy the other boundary condition
$u(\infty)=0$. If $m=1$, then the equations can be linearized about
the fixed point,
and $(1,0)$ is a hyperbolic saddle point with an unstable manifold in
the $(1,-1)$ direction and a stable manifold in the $(1,1)$ direction,
as shown in Fig.~\ref{fig:Lie}(a).

If $m\neq 1$, then the stable and unstable manifolds are degenerate at
linear order and form a cusp oriented in the $(0,1)$ direction, as
shown in Figs.~\ref{fig:Lie}(b) and (c). The nonlinear stability of the
fixed point can be determined by noting that $v^{\prime\prime} \sim
v^m$ as $u \rightarrow 1$. This equation has solutions satisfying
the boundary condition $v(-\infty) = 0$ if and only if $m \geq 1$.
A stable reaction front does not exist if $m<1$ because the
concentration of diffusing reactant A would become negative ahead of the
front.  Therefore, since $m \geq 1$ implies $\alpha = (m-1)/2(m+1)
\geq 0$,  the front width $w(t) \sim t^\alpha$ either stays the
same (for $m=1$) or increases (for $m > 1$) but cannot decrease in
time. For $m \geq 1$, we integrate $v^{\prime\prime} \sim v^m$ once
and substitute into (\ref{eq:vODE}) to obtain the separatrices in the
upper half-plane ($v>0$):
\begin{equation}
u \sim 1 \pm \sqrt{\frac{2}{m+1}} v^{(m+1)/2}, \ \mbox{as} \ (u,v)
\rightarrow (1,0^+) , \label{eq:separ}
\end{equation}
where the upper sign corresponds to the stable manifold ($\eta
\rightarrow \infty$) and the lower sign to the unstable manifold
($\eta \rightarrow -\infty$).

Let us briefly consider trajectories in the lower half-plane ($v <0$)
in the neighborhood of the fixed point ($u \approx 1$). Of course,
such trajectories are not physically allowed, but it is satisfying to
prove that the model equations exclude such possibilities.  If $m$ is
either irrational or a rational number of the irreducible form
$k_1/k_2$ where $k_2$ is even, then such trajectories do not exist
because in that case $v^m$ (and hence $u^\prime$) would not be a real
number. If $m = k_1/k_2$ where $k_1$ is even and $k_2$ is odd, then
the direction field $(u^\prime,v^\prime)$ is an even function of $v$,
which in light of (\ref{eq:separ}) implies that there are no other
separatrices in the lower half-plane and that trajectories merely
circle the fixed point, as shown in Fig.~\ref{fig:Lie}(b). Finally, if
$m=k_1/k_2$ where both $k_1$ and $k_2$ are odd, then $u^\prime$ is an
odd function of $v$ (while $v^\prime$ is even), and $(1,0)$ is a
saddle point.  In this case, Eq.~(\ref{eq:separ}) also describes
separatrices in the lower half-plane with the lower sign corresponding
to the stable manifold and the upper to the unstable manifold, as
shown in Fig.~\ref{fig:Lie}(c). As it leaves the fixed point, this
branch of the unstable manifold enters the region $(u>1,v<0)$,
throughout which $u^\prime>0$ and $v^\prime<0$, and thus it heads off
to $u=v=-\infty$ and cannot satisfy the other boundary condition, as
shown in Figs.~\ref{fig:Lie}(a) and (c). Therefore, any solutions must
lie entirely in the first quadrant of the phase plane ($u>0, v>0$).

In this way we are left with only one possible solution, which leaves
the fixed point along the unstable manifold of (\ref{eq:separ})
and enters the region defined by $0<u<1$ and $v>0$, throughout which
$u^\prime<0$ and $v^\prime>0$. Since the $v$-axis ($u=0$) is itself a
trajectory, which cannot be crossed, this candidate solution must
reach an asymptote $u(\infty)=u_o$, for some constant $0\leq u_o <
1$. However, it is clear from (\ref{eq:uODE}) that $u_o=0$ is the only
possible asymptote, which implies that the trajectory (if it exists)
must satisfy the other boundary condition $u(\infty)=0$. To check the
existence of this solution in the limit $s \rightarrow\infty$, note
that $v^\prime \sim 1$ which implies $u^\prime \sim -(s - s_o)^m u^n$
for some constant $s_o$ with $m>1$. Solutions to this equation
satisfying $u(\infty)=0$ exist if and only if $n \geq 1$. A stable
reaction front does not exist if $n < 1$ because the concentration of
static reactant B would be negative behind of the front.

\subsection{Concentration Profiles in the Reaction Front}

Although solutions to (\ref{eq:AcODE}) exist for $m,n \geq 1$, they
are not easily expressed in terms of elementary functions. The exact
trajectories in the phase plane, however, can be obtained. The ratio
of (\ref{eq:vODE}) and (\ref{eq:uODE}) yields a separable, first-order
equation for $v(u)$
\begin{equation}
\frac{dv}{du} = \frac{u-1}{v^m u^n} \label{eq:vu}
\end{equation}
which can be integrated to obtain the one-parameter family of
trajectories (for $u>0$) plotted in Fig.~\ref{fig:Lie}
\begin{equation}
(m+1)^{-1} v^{m+1} = 
\left\{
\begin{array}{ll}
c_1 + u - \log u     & \ \mbox{if} \ n=1 \\
c_2 +\log u + u^{-1} & \ \mbox{if} \ n=2 \\
c_n - (n-2) u^{2-n} + (n-1)^{-1} u^{1-n} & \ 
\mbox{if} \ n>1, n\neq 2 
\end{array}
\right.
\end{equation}
indexed by the real number $c_n$.

Applying the boundary conditions $v(u=1)=0$ (which determines $c_n$) and
$v(0)=\infty$ (which selects the positive branch when $v(u)$ is
multivalued), we arrive at the exact phase-plane trajectories (in the
region $0<u<1$, $v>0$) of the solution to the inner problem:
\begin{equation}
v = \left\{
\begin{array}{ll}
\left[ (m+1) \left(u - 1 - \log u\right) \right]^{1/(m+1)}
  & \ \mbox{if} \ n=1 \\
\left[ (m+1) \left(u^{-1} - 1 + \log u\right)\right]^{1/(m+1)} 
  & \ \mbox{if} \ n=2 \\
\left[\frac{(m+1)\left( 1 - (n-1)u^{2-n} +
(n-2)u^{1-n}\right)}{(n-1)(n-2)}\right]^{1/(m+1)} & \ \mbox{if} \ n>1,
n\neq 2
\end{array} \right.
\label{eq:Lie}
\end{equation}
which is an algebraic equation $v = g_{m,n}(u)$ relating $\Ac(\eta)$
and $\Bc(\eta)$ via (\ref{eq:uvdef}). This equation is transcendental,
but in some cases it is easily solved for $u = g_{m,n}^{-1}(v)$, {\it
e.g.} for $n=3$ we have
\begin{equation}
u = g_{m,3}^{-1}(v) = \frac{\sqrt{2(m+1)^{-1}v^{m+1}} - 1}
{2(m+1)^{-1}v^{m+1} - 1} .
\end{equation}
Note that $g_{m,n}^{-1}(0) = 1$ and $g_{m,n}^{-1}(\infty) = 0$.

By substituting (\ref{eq:Lie}) into (\ref{eq:vODE}) we arrive
at a first-order equation for $v(\eta)$
\begin{equation}
v^\prime = 1 - g_{m,n}^{-1}(v) \label{eq:vg}
\end{equation}
without any boundary conditions (because the conditions $v(-\infty)=0$
and $v^\prime(\infty)=1$ are automatically satisfied). Since
(\ref{eq:vg}) is separable, the solution to the inner problem can
be expressed in the form:
\begin{subeqnarray}
\Ac(\eta) & = & \Ac_1^{2/(m+1)} h_{m,n}^{-1}\left(
\Ac_1^{(m-1)/(m+1)}(\eta-\eta_o)\right) 
\slabel{eq:Ac_soln} \\ 
\Bc(\eta) & = & g_{m,n}^{-1}\left[h_{m,n}^{-1}\left(
\Ac_1^{(m-1)/(m+1)}(\eta-\eta_o)\right) 
\right]
\slabel{eq:Bc_soln}
\label{eq:AcBc_soln}
\end{subeqnarray}
where
\begin{equation}
h_{m,n}(v) \equiv \int_{v_o}^v \frac{ds}{1-g_{m,n}^{-1}(s)} ,
\label{eq:hmn}
\end{equation}
The precise location of the reaction front is set by choosing $v(s_o)
= v_o > 0$ for some constant $s_o = \Ac_1^{(m-1)/(m+1)}\eta_o$.  With
these results, the inner problem is reduced to the solution of two
algebraic equations for $g_{m,n}^{-1}(v)$ and $h_{m,n}^{-1}(\eta)$ and
one quadrature (\ref{eq:hmn}).  In practice, however, it is
simpler to integrate (\ref{eq:AcODE}) directly.

\begin{figure}[htb]
\begin{center}
\mbox{
\psfig{file=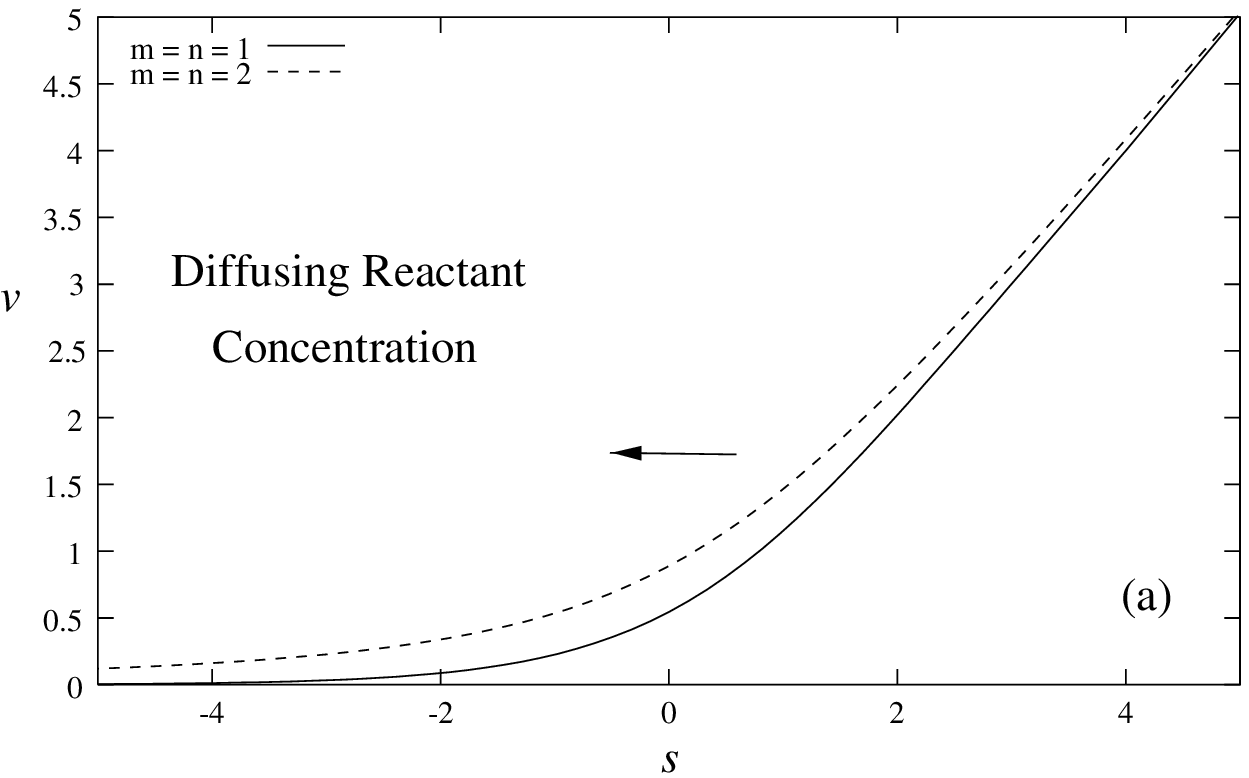,height=2.6in}
}
\mbox{
\psfig{file=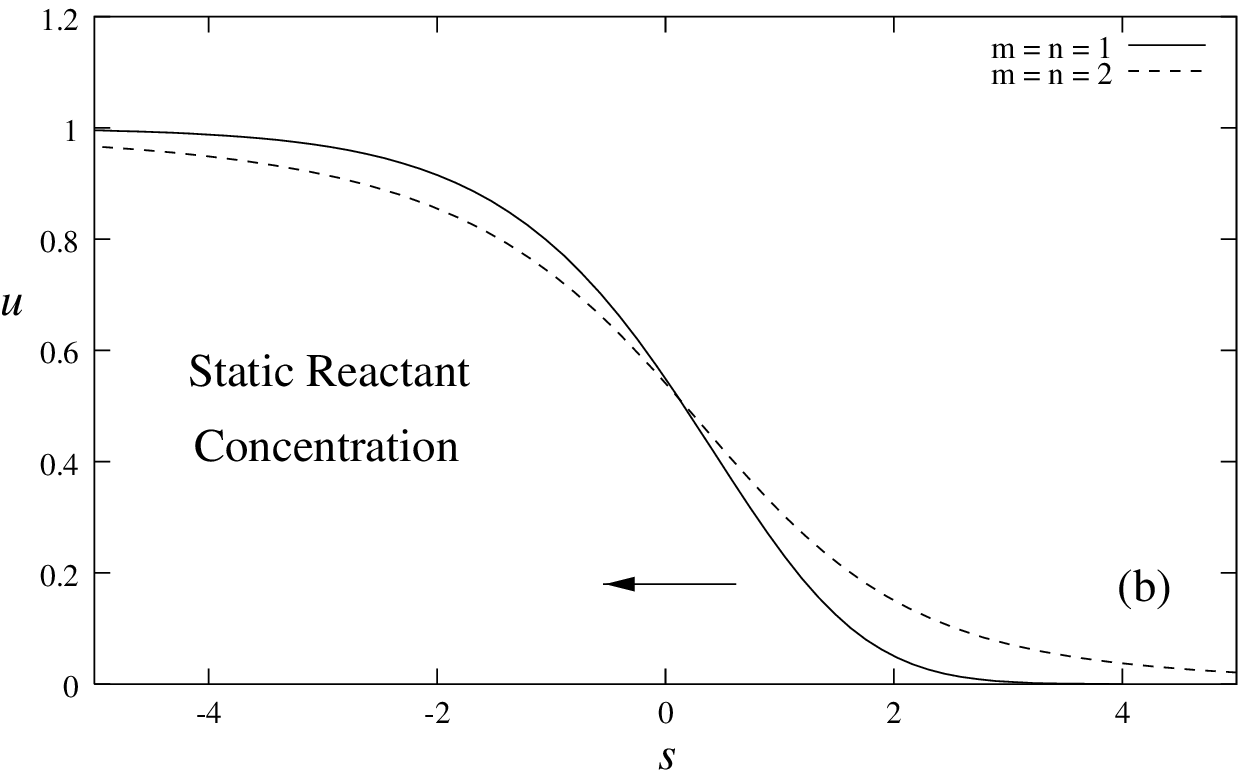,height=2.6in}
}
\mbox{
\psfig{file=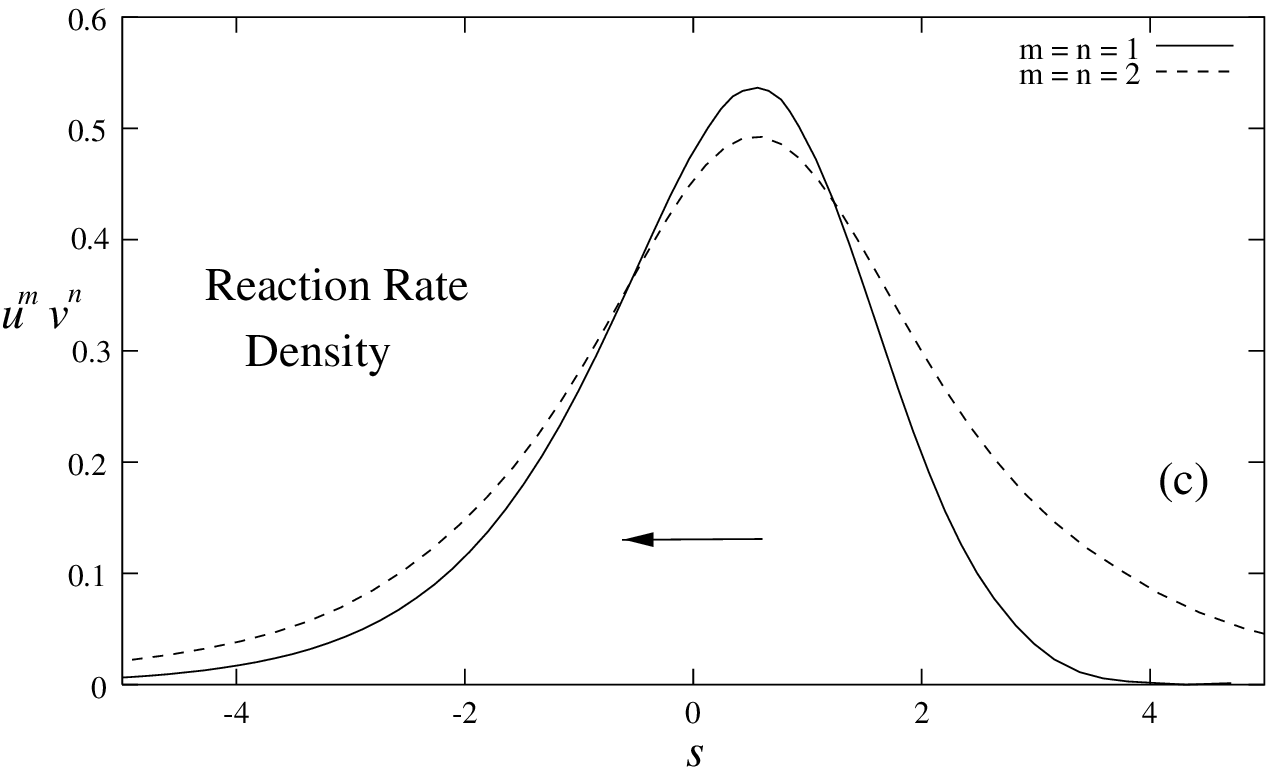,height=2.6in}
}
\begin{minipage}[h]{5.5in}
\caption{Structure of the reaction front obtained by numerical solutions of
Eq.~(\protect\ref{eq:veq}) with $s_o$ chosen to set $v(5)=5$. Profiles
of (a) $v(s) = \Ac(\eta) \Ac_1^{-2/(m+1)}$, (b) the static reactant
concentration $u(s) = \Bc(\eta)$ and (c) the reaction rate density
$u(s)^mv(s)^n = \Rc(\eta) \Ac_1^{-2m/(m+1)}$ versus $s = \eta \Ac_1^{(m-1)/(1+m)}$ are shown for
$m=n=1$ and $m=n=2$.  Arrows indicate that the reaction front
propagates from right to left.
\label{fig:Ac}
}
\end{minipage}
\end{center}
\end{figure}

Numerical solutions of the rescaled inner problem
\begin{equation}
v^{\prime\prime} = v^m(1-v^\prime)^n, \ \ \ v(-\infty)=0, \ \ \
v^{\prime}(\infty) = 1
\label{eq:veq}
\end{equation}
are obtained by a shooting method, with the results shown in
Fig.~\ref{fig:Ac}(a) for $m=n=1$ and $m=n=2$. The position of the
front is chosen such that $v(5)=5$. The static reactant concentration
$u(s)$ and the reaction rate density $v^m u^n$ are shown in
Figs.~\ref{fig:Ac}(b) and (c), respectively.  Note that the
concentration fields decay to their asymptotic values as $|\eta|
\rightarrow \infty$ more slowly as $m$ and $n$ are increased above unity, a
phenomenon that we explore analytically in next section.

\subsection{Localization of the Reaction Front}

The width of the reaction front varies in time according to $w(t) \sim
t^\alpha$ where $\alpha = (m-1)/2(m+1)$. On the scale of the diffusion
layer width $W(t) \sim t^{1/2}$, the reaction front is
\lq\lq localized" after long times because $\alpha < 1/2$. Note that
the overall localization of the front $w(t)/W(t)$ is controlled by $m$
(the reaction order of the diffusing species A), but we now show that
both $m$ and $n$ (the reaction order of the static species B) affect
localization on the scale of $w(t)$. Specifically, we derive the
spatial decay of the scaled reaction rate $\Rc(\eta) = \Ac(\eta)^m
\Bc(\eta)^n$ in terms of the inner similarity variable $\eta
\rightarrow \pm \infty$ (see Fig.~\ref{fig:Ac}(c)). The actual
reaction rate decays uniformly to zero in time, $r \sim
t^{-\beta}\Rc(\eta)$ with $\beta = m/(m+1)$, but here we are only
concerned with the shape of $\Rc(\eta)$.

Ahead of the reaction front in the limit $\eta \rightarrow -\infty$,
we have $\Ac^{\prime\prime} \sim \Ac^m$ from (\ref{eq:AcODE}), which
can be integrated to obtain the decay of concentration fields:
\begin{subeqnarray}
\Ac(\eta) & \sim & \left\{ \begin{array}{ll}
\Ac_2 e^{-|\eta|} & \ \mbox{if} \ m=1 \\
\Ac_2 |\eta|^{-2/(m-1)} & \ \mbox{if} \ m>1
\end{array} \right. \slabel{eq:Ac_ahead} \\
1- \Bc(\eta) & \sim & \left\{ \begin{array}{ll}
\Ac_3 e^{-|\eta|} & \ \mbox{if} \ m=1 \\
\Ac_3 |\eta|^{-(m+1)/(m-1)} & \ \mbox{if} \ m>1
\end{array} \right. \slabel{eq:Bc_ahead} \\
\Rc(\eta) & \sim & \left\{ \begin{array}{ll}
\Ac_2^m e^{-m|\eta|} & \ \mbox{if} \ m=1 \\
\Ac_2^m |\eta|^{-2m/(m-1)} & \ \mbox{if} \ m>1
\end{array} \right.
\end{subeqnarray}
where $\Ac_2$ and $\Ac_3$ are constants. Note that the localization of
$\Ac(\eta)$ and $\Rc(\eta)$ ahead of the front is entirely controlled
by the reaction order $m$ of the depleted reactant (which is the
diffusing species $A$). There is a transition from an exponential
decay for $m=1$ to a slower power-law decay for $m>1$.

Next we consider localization of $\Rc(\eta)$ behind the reaction front
in the limit $\eta \rightarrow \infty$. From asymptotic matching with
the diffusion layer we have already derived $\Ac(\eta) \sim \Ac_1
\eta$. The asymptotic decay of $\Bc(\eta)$ and $\Rc(\eta)$ is 
obtained by integrating $\Ac_1 \Bc^\prime \sim - (\Ac_1 \eta)^m
\Bc^n$:
\begin{subeqnarray}
\slabel{eq:Bctail}
\Bc(\eta) & \sim & \left\{ \begin{array}{ll}
\Bc_1 \exp\left[-\Ac_1^{m-1}\eta^{m+1}/(m+1)\right] & \ \mbox{if} \ n=1 \\
\Bc_1 \eta^{-(m+1)/(n-1)} & \ \mbox{if} \ n>1
\end{array} \right. \\
\slabel{eq:Rctail}
\Rc(\eta) & \sim & \left\{ \begin{array}{ll}
\Ac_1^m \Bc_1^n \eta^m \exp\left[-\Ac_1^{m-1}\eta^{m+1}/(m+1)\right]
        & \ \mbox{if} \ n=1 \\
\Ac_1^m \Bc_1^n \eta^{-(m+n)/(n-1)} & \ \mbox{if} \ n>1
\end{array} \right.
\end{subeqnarray}
where $\Bc_1$ is a constant. Once again, a higher reaction order $n$
for the depleted species (which is the static species B) broadens the
front: There is another transition from exponential decay for $n=1$ to
a power-law decay for $n>1$.  Note, however, that increasing the
reaction order $m$ of the diffusing species A contracts the back side
of the front.

The fact that $\Rc(\eta)$ has a fairly broad, power-law decay for $n >
1$ toward the diffusion layer should cause concern since we have
previously assumed in (\ref{eq:Rzero}) that the reaction term is
negligible in the diffusion layer. Fortunately, however, for all
$m,n\geq 1$ the decay of $\Bc(\eta)$ is just fast enough to satisfy
(\ref{eq:Rzero}) in the intermediate region where $B(0<\zeta\ll 1)
\approx \Bc(\eta \gg 1)$. From (\ref{eq:Bctail}) along with
$\tilde{A}(\zeta,t) \sim A(\zeta)$, $\tilde{\Bc}(\eta,t) \sim
\Bc(\eta)$, $\eta = 2\zeta t^{1/(m+1)}$ and $\delta = 1/2$, we have
\begin{equation}
t^{2\delta} \tilde{A}(\zeta,t)^m \tilde{B}(\zeta,t)^n =
\left\{ \begin{array}{ll}
O\left( t \exp\left[-\Ac_1^{m-1}(2\zeta)^{m+1}t/(m+1)\right] \right)
        & \ \mbox{if} \ n = 1 \\
O\left( \zeta^{-n(m+1)/(n-1)} t^{-n/(n-1)} \right)
        & \ \mbox{if} \ n > 1
\end{array}
\right. 
\end{equation}
as $t \rightarrow \infty$ with $0<\zeta\ll 1$ fixed, which verifies 
(\ref{eq:Rzero}) in the intermediate region for any $m, n \geq
1$. However, we now prove that (\ref{eq:Rzero}) actually holds
throughout the diffusion layer for all $\zeta > 0$.

\section{Transient Decay in the Diffusion Layer}
\label{sec:transients}

\subsection{Assumption of Quasi-Stationarity}

The analysis of long-time asymptotics in the previous section rests on
two basic assumptions: $(i)$ \lq\lq scale separation", given by
\begin{subeqnarray}
0 < \lim_{t\rightarrow \infty} \tilde{\Ac}(\eta,t)^m
\tilde{B}(\zeta,t)^n  & < &  \infty \ \ \ \mbox{for} \ |\eta| < \infty \\
\lim_{t\rightarrow \infty} t \cdot \tilde{A}(\zeta,t)^m
\tilde{B}(\zeta,t)^n & = &  0 \ \ \ \mbox{for} \ \zeta > 0 ,
\slabel{eq:SS}
\label{eq:SSboth}
\end{subeqnarray}
and $(ii)$ \lq\lq quasi-stationarity'', given by (\ref{eq:AcBc_converge})
and (\ref{eq:AB_converge}) along with  
\begin{subeqnarray}
\lim_{t\rightarrow \infty} t^{(m-1)/(m+1)}\frac{\partial
\tilde{\Ac}}{\partial t}(\eta,t) & = & 0 \ \ \ \mbox{for} \
|\eta|<\infty \slabel{eq:QSAc} \\
\lim_{t\rightarrow \infty} t^{m/(m+1)}\frac{\partial
\tilde{\Bc}}{\partial t}(\eta,t) & = & 0 \ \ \ \mbox{for} \
|\eta|<\infty \slabel{eq:QSBc} \\
\lim_{t\rightarrow \infty} t \cdot \frac{\partial
\tilde{A}}{\partial t}(\zeta,t) & = & 0 \ \ \ \mbox{for} \
\zeta > 0 \slabel{eq:QSA}  .
\label{eq:QS}
\end{subeqnarray}
Assumption $(i)$ states that two spatial regions with disparate length
scales, the reaction front and the diffusion layer, arise where the
reaction term on the right-hand side of the governing partial
integro-differential equation (\ref{eq:ide1}) or (\ref{eq:ide2}) is,
respectively, either comparable to or dominated by the diffusion
term. Assumption $(ii)$ states that, when viewed on scales appropriate
for each region, the solution to the initial-boundary-value problem
(\ref{eq:eqs}) approaches an asymptotically self-similar form, which
is suggested by the fact that there is no natural length scale in the
problem.

These ubiquitous assumptions~\cite{galfi,koza1,koza2} have been
rigorously justified~\cite{schenkel,vanbaalen} in the special case of
a perfectly symmetric ($D_A=D_B$, $\rho_A^o=\rho_B^o$), and thus
stationary ($\nu=0$), reaction front involving two diffusing reactants
with certain kinetic orders ($m=n=1$ and $m=n>3$).  To our knowledge,
a similar mathematical validation of these assumptions has not been
given for the general situation of a moving reaction front with
arbitrary kinetic orders for either one or two diffusing reactants.
For one static reactant, however, convergence at the diffusive scale
has been rigorously established by Hilhorst et al.~\cite{peletier96},
even with a very general reaction term~\cite{peletier99}.

In this section, we prove the more modest result that
quasi-stationarity implies scale separation, {\it i.e.}  (\ref{eq:QS})
implies (\ref{eq:SSboth}). (It suffices to show (\ref{eq:SS}) since
(\ref{eq:SSboth}) follows from the definition of $\eta$ in
section~\ref{sec:Ac}.)
Although this analysis justifies {\it a posteriori} the assumption
of scale separation in our fairly general situation ($D_B=0$, $\nu >
0$, $m,n \geq 1$), it more importantly reveals the transient decay
to the asymptotic similarity solution in the diffusion
layer. Specifically, we derive exact formulae for the asymptotic
decay of the reaction-rate density and static-reactant concentration
in the diffusion layer.

We begin by precisely stating our assumptions related to
(\ref{eq:QS}).  From (\ref{eq:QSAc}) and its consequence
(\ref{eq:Ac_ahead}), we conclude that the diffusing reactant
concentration vanishes on the scale $W(t) = \sqrt{t}$ ahead of the
front (where there is no diffusion layer) since $\tilde{A}(\zeta,t)
\sim \Ac(2\zeta t^{1/(m+1)}) \rightarrow 0$ as $t \rightarrow \infty$
with $\zeta < 0$ fixed. This result can be combined with
(\ref{eq:A_converge}) to obtain a statement of quasi-stationarity on
the scale $W(t)$:
\begin{equation}
\tilde{A}(\zeta,t) \rightarrow  A(\zeta)H(\zeta) \ \ \mbox{and} \ \ 
\frac{\partial \tilde{A}}{\partial \zeta}(\zeta,t) \rightarrow 
A^\prime(\zeta)H(\zeta) 
\label{eq:AH} 
\end{equation}
as $t \rightarrow \infty$ with $\zeta \neq 0$ fixed.  We have already
derived the exact form of the similarity function $A(\zeta)$ in
(\ref{eq:diffA}) as a consequence of neglecting reactions
(\ref{eq:SS}). To avoid a circular argument, however, we must now
establish (\ref{eq:SS}) without using (\ref{eq:diffA}), thus giving
{\it a posteriori} justification for the latter equation. Throughout
section~\ref{sec:transients} our only assumptions about $A(\zeta)$ are
$A(0)=0$ and $A^\prime(0)>0$. These properties follow from matching
with the reaction front, where $a \rightarrow 0$ ({\it i.e.} $\gamma >
0$) follows from quasi-stationarity, as shown in section~\ref{sec:Ac}.

\subsection{Direction of the Diffusing-Reactant Flux}

Let us prove that $A(\zeta)$ is strictly increasing in the diffusion
layer, $A^\prime(\zeta) > 0$ for all $\zeta \geq 0$, as a consequence
of (\ref{eq:SSboth}).  Combining (\ref{eq:QSA}) with (\ref{eq:Along}),
we have
\begin{equation}
-2(\zeta-\nu)\frac{\partial \tilde{A}}{\partial \zeta} \sim
\frac{\partial^2 \tilde{A}}{\partial \zeta^2} - t \tilde{A}^m
\tilde{B}^n  \ \ \ \mbox{as} \ t\rightarrow \infty \ \mbox{with} \
\zeta > 0 \ \mbox{fixed}
\end{equation}
which is easily integrated once using an integrating factor,
\begin{equation}
\frac{\partial \tilde{A}}{\partial \zeta}(\zeta,t) 
  \sim e^{-(\zeta-\nu)^2} \left[ e^{\nu^2} \frac{\partial
  \tilde{A}}{\partial \zeta}(0,t) + t \int_0^\zeta \tilde{A}(\xi,t)^m
  \tilde{B}(\xi,t)^n e^{(\xi-\nu)^2}d\xi
\right] .
\label{eq:dAt}
\end{equation}
Since the second term on the right-hand side 
is non-negative for all $t>0$, we can pass to the limit $t\rightarrow
\infty$ for any fixed $\zeta> 0$ to obtain the desired bound
\begin{equation}
A^\prime(\zeta) \geq A^\prime(0)e^{\nu^2-(\zeta-\nu)^2} > 0 ,
\label{eq:dAbound}
\end{equation}
which expresses the physical fact that everywhere in the diffusion
layer a nonzero flux of the diffusing species is directed toward the
reaction front (at sufficiently large times).

\subsection{Decay of the Static-Reactant Concentration}

We now prove that $b(x,t)$ vanishes asymptotically in the diffusion
layer as a consequence of quasi-stationarity, which implies
\begin{equation}
\tilde{B}(\zeta,t) \rightarrow B(\zeta) = H(-\zeta)\ \ \ \mbox{as} \ t
\rightarrow \infty  
\ \mbox{with} \ \zeta \neq 0 \ \mbox{fixed} . \label{eq:Bzeta}
\end{equation}
For $\zeta < 0$, this follows from (\ref{eq:Bc_ahead}) since there is
no diffusion layer ahead of the front, and therefore
$\tilde{B}(\zeta,t) \sim \Bc(2\zeta t^{1/(m+1)}) \rightarrow 1 =
B(\zeta)$ as $t \rightarrow \infty$ with $\zeta < 0$ fixed.  Likewise,
in section \ref{sec:dim}, we have already established (\ref{eq:Bzeta})
for $\zeta > \nu$ since $\tilde{B}(\zeta,t)$ vanishes there
identically for all times. Therefore, it only remains to prove that
$B(\zeta) = 0$ for $0 < \zeta < \nu$.

In light of the expression for $b(x,t)$ in (\ref{eq:bint}), the
definition of $\phi_m(x,t)$ in (\ref{eq:phidef}) and the restriction
$n \geq 1$, it suffices to show that $\Phi_m(\zeta,t) \equiv
\phi_m(x,t) \rightarrow \infty$ as $t \rightarrow \infty$ for  $0 <
\zeta < \nu$  fixed. Using $\zeta = \nu + x/2\sqrt{t}$ (since
$\delta=\sigma=\frac{1}{2}$), we transform $\phi_m(x,t)$ into the
diffusion-layer coordinates $(x,t) \mapsto (\zeta,t)$
\begin{equation}
\Phi_m(\zeta,t) =  \int_0^t a\left(2\sqrt{t}(\zeta-\nu),\tau\right)^m
d\tau ,
\end{equation}
and express this in terms of the diffusion-layer scaling function
$a(x,\tau) = \tilde{A}(\nu + x/2\sqrt{\tau},\tau)$
\begin{equation}
\Phi_m(\zeta,t) = \int_0^t \tilde{A}\left(\sqrt{\frac{t}{\tau}}
(\zeta-\nu)+\nu, \tau\right)^m d\tau .
\end{equation}
It is convenient to work with the partial time-derivative of
$\Phi_m(\zeta,t)$ given by the Leibniz rule:
\begin{equation}
\frac{\partial \Phi_m}{\partial t} =  \tilde{A}(\zeta,t)^m +
\int_0^t \frac{\partial \tilde{A}^m}{\partial \zeta}
\left(\sqrt{\frac{t}{\tau}}(\zeta-\nu)+\nu, \tau\right)
\frac{1}{2t}\sqrt{\frac{t}{\tau}}(\zeta-\nu)d\tau .
\end{equation}
Focusing on the region $0 < \zeta < \nu$, we make the transformation $\xi
= \sqrt{t/\tau}(\zeta - \nu) + \nu$,
\begin{equation}
\frac{\partial \Phi_m}{\partial t} =  \tilde{A}(\zeta,t)^m 
- \int_{-\infty}^\zeta
\frac{\partial \tilde{A}^m}{\partial\zeta}
\left(\xi,\left(\frac{\zeta-\nu}{\xi-\nu}\right)^2t\right)
\left(\frac{\zeta-\nu}{\xi-\nu}\right)^2 d\xi  ,
\label{eq:phitrans}
\end{equation}
and pass the limit $t\rightarrow \infty$ inside the integral to obtain
\begin{equation}
\lim_{t\rightarrow \infty}\frac{\partial \Phi_m}{\partial t} = 
A(\zeta)^m  - \int_0^\zeta \frac{dA^m}{d\zeta}(\xi)
\left(\frac{\zeta-\nu}{\xi-\nu}\right)^2 d\xi ,
\label{eq:dphi0}
\end{equation}
where the lower limit of integration follows from (\ref{eq:AH}) since
$A(\zeta) = A^\prime(\zeta) = 0$ for $\zeta < 0$. This step is
justified by the Dominated Convergence Theorem~\cite{evans} because,
by virtue of (\ref{eq:AH}), there exist constants $M, t_o > 0$ such
that the integrand in (\ref{eq:phitrans}) is bounded for all $t > t_o$
by $M/(\xi - \nu)^2$, which is integrable on $(-\infty,\zeta)$, if
$\zeta < \nu$.

Since $A(0) = 0$ is required by matching between the two regions of
quasi-stationarity, Eq.~(\ref{eq:dphi0}) can be written in the form
$\partial \Phi_m/\partial t \sim f_m(\zeta)$, where
\begin{equation}
f_m(\zeta) \equiv \int_0^\zeta
\frac{dA^m}{d\zeta}(\xi) \left[ 1 -
\left(\frac{\zeta-\nu}{\xi-\nu}\right)^2\right] d\xi . 
\label{eq:fm}
\end{equation}
Note that $f_m(\zeta)>0$ for $0 < \zeta < \nu$ since
$A^\prime(\zeta)>0$ in this region, as shown in
(\ref{eq:dAbound}). Therefore, with an integration of (\ref{eq:fm}),
we arrive at the desired result
\begin{equation}
\Phi_m(\zeta) \sim f_m(\zeta)t
\ \ \mbox{as} \ \ t \rightarrow \infty \ \ \mbox{with} 
\ \ 0 < \zeta < \nu \ \ \mbox{fixed,}
\label{eq:Phit}
\end{equation}
thus completing the proof that $B(\zeta) = 0$ for $\zeta> 0$.

By substituting (\ref{eq:Phit}) into (\ref{eq:bint}), we obtain the
transient decay of $\tilde{B}(\zeta,t)$ in part of the diffusion
layer where the reaction front has already passed ($0 < \zeta \leq
\nu$):
\begin{equation}
\tilde{B}(\zeta,t) \sim \left\{ \begin{array}{ll}
e^{-qf_m(\zeta)t} & \ \mbox{if} \ n=1 \\
\left[ q(n-1)f_m(\zeta) t \right]^{-1/(n-1)}  & \
\mbox{if} \ n >1
\end{array} \right. \label{eq:Blimit}
\end{equation}
Note that $\tilde{B}(\zeta,t)$ vanishes with exponential decay if
$n=1$ and with a power-law decay if $n > 1$. Therefore, by measuring
the asymptotic decay (either exponential or power-law) of the
static-reactant concentration in the diffusion layer, the reaction
order $n$ could in principle be inferred from experimental data
(although such measurements are difficult in practice~\cite{leger}).

\subsection{Decay of the Reaction-Rate Density}

From (\ref{eq:Blimit}) we easily obtain the asymptotic decay on the
reaction rate density in the diffusion layer as $t \rightarrow \infty$
with $\zeta>0$ fixed:
\begin{equation}
t \tilde{A}(\zeta,t)^m \tilde{B}(\zeta,t)^n \sim \left\{
\begin{array}{ll} A(\zeta)^m t e^{-qf_m(\zeta)t} & \ \mbox{if} \ n=1 \\
A(\zeta)^m \left[ q(n-1)f_m(\zeta)\right]^{-\frac{n}{n-1}}
t^{-\frac{1}{n-1}} & \
\mbox{if} \ n >1
\end{array} \right.  ,
\label{eq:Rtransient}
\end{equation}
which establishes (\ref{eq:SS}). The reaction term in the diffusion
layer has previously been neglected based only on physical
intuition~\cite{koza2}, but here we have given a mathematical
justification.

\subsection{The Decay Time When $n=1$}

\begin{figure}[htb]
\begin{center}
\mbox{
\psfig{file=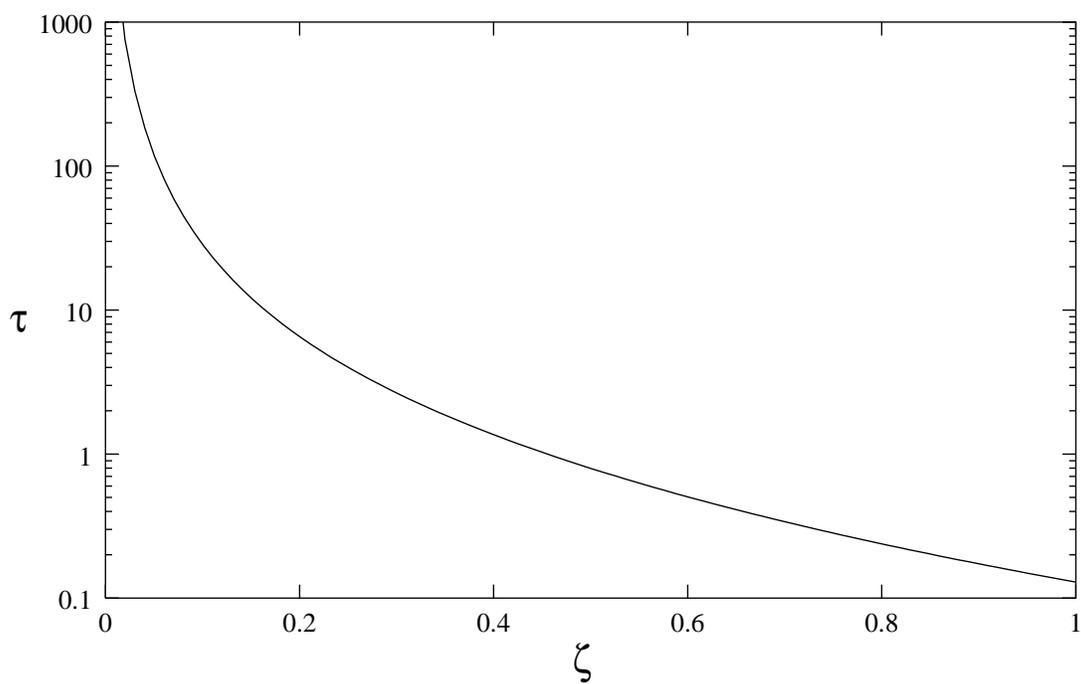,height=3.5in}
}
\begin{minipage}[h]{5.5in}
\caption{Transient decay time $\tau_1(\zeta)$ given by
(\protect\ref{eq:tau}) for the reaction rate and static-reactant
concentration in the diffusion layer when $m=n=1$.
\label{fig:tau}
}
\end{minipage}
\end{center}
\end{figure}

Since the reaction term vanishes sufficiently fast in the diffusion
layer to justify {\it a posteriori} the analysis in
section~\ref{sec:similarity}, the exact expression for $A(\zeta)$ from
(\ref{eq:diffA}) may be substituted into (\ref{eq:fm}) to
evaluate the function $f_m(\zeta)$. If $n>1$, then $f_m(\zeta)$
affects the transient decay in (\ref{eq:Blimit}) and
(\ref{eq:Rtransient}) only as a multiplicative prefactor in a
power law, which would be difficult to measure in a real experiment.
If $n=1$, however, then $f_m(\zeta)$ sets the characteristic time
$\tau_m(\zeta)^{-1} \equiv q f_m(\zeta)$ of an exponential decay, which
is easier to measure experimentally. Therefore, we
now derive an exact expression for the decay time $\tau_1(\zeta)$
($0<\zeta\leq \nu$) in the case $m=1$:
\begin{eqnarray}
\tau_1(\zeta)^{-1} & = & q f_1(\zeta) \nonumber \\
& = & \int_0^\zeta 2\nu e^{\nu^2-(\nu-\xi)^2} \left[ 1 -
\left(\frac{\zeta-\nu}{\xi-\nu}\right)^2\right] d\xi \nonumber \\
& = & 2(\nu-\zeta)\left[ (\nu-\zeta) - \nu
e^{\nu^2-(\nu-\zeta)^2}\right] \nonumber \\
& & + \sqrt{\pi} \nu e^{\nu^2}
\left[ 1 + 2(\nu-\zeta)^2 \right]\cdot \left[ \erf(\nu) -
\erf(\nu-\zeta) \right] . \label{eq:tau}
\end{eqnarray}
Note that $\tau_1(0) = \infty$ in the vicinity of the reaction front
($\zeta = 0$) because (\ref{eq:Rtransient}) and (\ref{eq:Blimit}) no
longer hold.  Within the diffusion layer, the decay time is a
decreasing function of distance $\zeta$ away from the reaction front,
as shown in Fig.~\ref{fig:tau}.
These results may be used to infer reaction orders and perhaps even
kinetic constants in diffusion-limited corrosion experiments from
transient decay measurements of the reaction-rate density in the
diffusion layer~\cite{leger}.

\section{Uniformly Valid Asymptotic Approximations}
\label{sec:uniform}

In the previous two sections we have argued for the existence of a
unique asymptotic similarity solution (up to an unknown constant
$\eta_o$) contingent upon certain \lq\lq quasi-stationarity"
conditions, which are likely to be satisfied for the specified initial
conditions (see below). This solution, valid after long times,
consists of two different asymptotic approximations for $a(x,t)$, the
concentration of the diffusing reactant A, which reflect the different
couplings of length and time in the reaction front and the diffusion
layer. A single asymptotic approximation for $a(x,t)$ that is
uniformly valid across all space is obtained by adding the two
contributions from the reaction front (the inner region) and the
diffusion layer (the outer region) and subtracting the overlap (from
the intermediate region)~\cite{hinch,bender}:
\begin{equation}
a(x,t) \sim \left[\Ac(\eta-\eta_o)- \Ac_1 \cdot (\eta-\eta_o)
H(\eta-\eta_o)\right]t^{-1/(m+1)} + A(\zeta) H(\zeta), \ \mbox{as} \ t
\rightarrow\infty \ \mbox{for\ all\ } x . \label{eq:a_unif}
\end{equation}
where the reaction-front and diffusion-layer similarity variables are
\begin{subeqnarray}
\eta(x,t) & = &  \frac{x + 2\nu t^{1/2}}{t^{(m-1)/2(m+1)}}  \\
\mbox{and} \ \ \zeta(x,t) & = &  \frac{x + 2\nu t^{1/2}}{2t^{1/2}} ,
\end{subeqnarray}
$\nu(q)^2$ is  the diffusion constant of the reaction 
front (see (\ref{eq:qnu}) and Fig.~\ref{fig:nuq}), $\Ac_1 =
\nu(q)/q$ is a constant proportional to the diffusive flux entering
the front, $\Ac(\eta)$ is the reaction-front similarity function (see
(\ref{eq:Ac_soln}) and Fig.~\ref{fig:Ac}), $A(\zeta)$ is the
diffusion-layer similarity function (see (\ref{eq:diffA}) and
Fig.~\ref{fig:A}) and $\eta_o$ is an undetermined constant depending
upon the initial conditions as well as the precise definition of the
reaction-front location. The uniform approximation has been determined
analytically up to the solution of two algebraic equations
(\ref{eq:AcBc_soln}) and one quadrature (\ref{eq:hmn}).

A subtle point in the construction of this uniformly valid
approximation is that shifting the position of the front by $\eta
\mapsto \eta - \eta_o$ does not affect matching with the diffusion
layer because in that case $\zeta \mapsto (\eta-\eta_o)t^{-1/(m+1)}/2
\sim \eta t^{-1/(m+1)}/2 = \zeta$. In other words, because the
reaction front is \lq\lq infinitely thin" compared to the diffusion
layer, translating its similarity variable by a constant $\eta_o$, or
any other function of time that is $o(t^{1/2})$, does not require that
the diffusion-layer similarity variable $\zeta$ be shifted as well.

The situation for $b(x,t)$, the concentration of the static reactant
B, is much simpler. By comparing the asymptotic bound on $b(x,t)$ in
the diffusion layer given by (\ref{eq:Blimit}) with the tail of
the reaction-front approximation given by (\ref{eq:Bctail}) with
$\eta = 2\zeta t^{1/(m+1)}$, we see that that the asymptotic behavior
of $b(x,t)$ is identical in the two regions. Therefore,
\begin{equation}
b(x,t) \sim \Bc(\eta-\eta_o), \ \mbox{as}
\ t \rightarrow\infty \ \mbox{for\ all\ } x \label{eq:b_unif}
\end{equation}
is a uniformly valid approximation, where $\Bc(\eta)$ is the
reaction-front similarity function given by (\ref{eq:Bc_soln}).

At this point the initial conditions have not yet entered the
analysis except in (\ref{eq:bint}), which only influences the
prefactors of the transient-decay formulae in
section~\ref{sec:transients}. Therefore, the asymptotic similarity
solution is universal up to a constant shift of the reaction front by
$\eta_o$ for some broad set of initial conditions which presumably
contains (\ref{eq:ic}). In general, this \lq\lq universality class''
of initial conditions leading to the same asymptotic similarity
solution (up to different values of $\eta_o$) is expected to be
attained whenever the initial reaction-rate distribution $r(x,0) =
a(x,0)^m b(x,0)^n$ is sufficiently well localized and the reactants
are sufficiently well separated. This class surely contains all
initial conditions for which $r(x,0)$ has compact support, {\it e.g.}
$r(x,0) = 0$ for $x\neq 0$ in (\ref{eq:ic}), or exponential decay,
{\it e.g.} $r(x,0) < Me^{-|x|/x_o}$ for some $M,x_o>0$, but perhaps
not slower power-law decay.

\section{Discussion}
\label{sec:disc}

In this article we have studied the long-time asymptotics of solutions
to the initial-boundary-value problem of
(\ref{eq:rhoBC})--(\ref{eq:rho}), which is a generic mean-field model
for the corrosion of a porous solid by a diffusing chemical. We have
derived a uniformly valid asymptotic approximation
(\ref{eq:a_unif})--(\ref{eq:b_unif}) consisting of matched similarity
solutions in two distinct regions, the reaction front and diffusion
layer, each possessing different power-law scaling behavior.  The
existence and uniqueness of the similarity functions and the scaling
exponents have been established if and only if $m,n \geq 1$, and
through an analysis of transients in the diffusion layer the
asymptotic scale separation has been shown to follow from the
assumption of quasi-stationarity. Since quasi-stationarity has been
observed in recent experiments on the corrosion of ramified
electrodeposits~\cite{leger}, the present analysis therefore suffices
to establish the theoretical predictions of the mean-field equations
for at least one particular corrosion system. Although the case
considered here ($m,n\geq 1$, $q \neq 1$) is more complicated, it
would be useful to perform a rigorous transient analysis along the
lines of Schenkel {\it et al.}~\cite{schenkel} (who considered only
the case $D_A=D_B\neq 0$, $q=1$ and $m=n=1$). Nevertheless, we have at
least provided a firm mathematical justification for the scale
separation between the diffusion layer and reaction front.

In this work we have paid special attention to the effect of
higher-order reactions ($m,n > 1$). First of all, the scaling
exponents vary with the reaction order $m$ of the diffusing reactant
in precisely the same way as they do on the sum $m+n$ in the case of
two diffusing reactants~\cite{cornell}, as shown in Table
~\ref{table:exponents}. Moreover, the spatial localization of the
reaction rate $r(a,b)$ on each side of the front depends primarily on
the reaction order of the depleted reactant: As the appropriate
reaction order is increased from unity, the spatial dependence of the
reaction rate away from the front changes from an exponential decay to
a progressively broader power-law decay. Similarly, the temporal decay
of the depleted (static) reactant concentration in the diffusion layer
depends sensitively on its reaction order, undergoing a 
transition from exponential to power-law decay (in time) as $n$ is
increased from unity.  These properties may have general relevance for
more complicated multi-component reaction-diffusion systems.

Other qualitiative features of our analysis that might have more
general applicability are the dominant balances in the reaction front,
where the concentration of a diffusing reactant is determined by a
balance between reactions and \lq\lq steady state" diffusion, {\it
i.e.} a mobile species diffuses slowly to the front where it
immediately reacts. On the other hand, the concentration of a static
reactant is determined by a balance between reactions and fictitous
advection due to the moving reference frame, {\it i.e.} since the
static species cannot diffuse to the front, the front must diffuse to
it. These guiding principles might help simplify more complicated
reaction-diffusion equations for which asymptotically self-similar
solutions do not exist.

\section*{Epilogue}

After the writing of this article, a referee pointed out some
interesting similarities between our analysis of chemical reaction
fronts and various existing studies in combustion
theory~\cite{ZFK,zeldovich}. Indeed, the equations for combustion
waves introduced by Zeldovich and Frank-Kamenetzki~\cite{ZFK}, which
have since been generalized by many authors (e.g. Matkowsky and
Sivashinsky~\cite{MS}), bear some resemblance to (\ref{eq:coup}) since
they describe the diffusion of a fuel substance coupled to the
diffusion of heat. In combustion theory, however, the usual reaction
term is quite different from (\ref{eq:Rpower}) because it involves
exponential Arrhenius temperature dependence, and the initial and
boundary conditions also differ from those considered here.  As a
result, combustion waves tend to exhibit qualitatively different
behavior from reaction-diffusion fronts. (For example, simple flame
fronts have constant width and constant velocity.)  Nevertheless,
combustion waves exhibit multiple scales analogous to the
reaction-front and diffusion layers described here, which have also
been analyzed using matched asymptotic expansions~\cite{schult}
(although not in the dynamical setting of this work). The idea of
matching derivatives between the inner and outer regions actually
appears to have its origin in the pioneering paper of Zeldovich and
Frank-Kamenetzki~\cite{ZFK}, in which the velocity of a simple flame
front is determined by analyzing a single-component equation like
(\ref{eq:single}).  In hindsight, it is somewhat surprising that the
recent parallel literatures on two-species reaction-diffusion fronts
and combustion waves have developed quite independently of each other,
without any cross-references (at least, none to our knowledge). It is
hoped, therefore, that this paper will initiate the ``diffusion'' of
ideas between these two mature but related disciplines.

\section*{Acknowledgments}

The authors thank C. L\'eger and R. R. Rosales for useful discussions.
This work was supported by an NSF infrastructure grant (MZB) and
grants from the Harvard MRSEC DMR-980-9363 and the Army Research
Office DAAG-55-97-1-0114 (HAS).

\begin{table}[htb]
\begin{center}
\begin{tabular}{lccc}
        & $\alpha$ & $\beta$ &  $\gamma$
\\
\\
$(i) \ \ \ D_A > 0,\ D_B = 0$
        & $\frac{m-1}{2(m+1)}$ &$\frac{m}{m+1}$
        &  $\frac{1}{m+1}$
\\
\\
$(ii) \ \ D_A > 0,\ D_B > 0$
        & $\frac{m+n-1}{2(m+n+1)}$ & $\frac{m+n}{m+n+1}$
        & $\frac{1}{m+n+1}$
\\
\\
\end{tabular}
\begin{minipage}[h]{6in}
\caption{Comparison of the two possible sets of scaling exponents
for asymptotic similarity solutions to the one-dimensional, mean-field
reaction-diffusion equations for two initially separated
reactants. The width of the reaction front is given by $w \sim
t^\alpha$, the magnitude of the reaction rate by $R \sim t^{-\beta}$
and the concentration of a diffusing species in the reaction front by
$a \sim t^{-\gamma}$.
\label{table:exponents}
}
\end{minipage}
\end{center}
\end{table}

\end{document}